\documentclass[sigconf]{acmart}

\usepackage[utf8]{inputenc}
\usepackage{url}
\usepackage{amsfonts}
\usepackage{nicefrac}
\usepackage{algorithm}
\usepackage[noend]{algpseudocode}
\usepackage{algorithmicx}
\usepackage{subfigure}
\usepackage{amsmath}
\usepackage{amsthm}
\usepackage{xcolor,soul}
\usepackage{tabularx}
\usepackage{verbatim}
\usepackage{flushend}
\usepackage[multiple]{footmisc}
\usepackage{capt-of}
\usepackage[normalem]{ulem}
\usepackage{wasysym}

\newcommand{\eat}[1]{}

\theoremstyle{definition}

\setcopyright{none}
\begin{document}
    \title{M6-Rec: Generative Pretrained Language Models are Open-Ended Recommender Systems}

    \author{Zeyu Cui$^*$, Jianxin Ma$^*$, Chang Zhou, Jingren Zhou, Hongxia Yang}
    \email{{zeyu.czy, jason.mjx, ericzhou.zc, jingren.zhou, yang.yhx}@alibaba-inc.com}
    \affiliation{%
        \institution{DAMO Academy, Alibaba Group}
        \city{}
        \country{}
    }

    \renewcommand{\shortauthors}{Cui et al.}

    \begin{abstract}
        Industrial recommender systems have been growing increasingly complex, may involve \emph{diverse domains} such as e-commerce products and user-generated contents, and can comprise \emph{a myriad of tasks} such as retrieval, ranking, explanation generation, and even AI-assisted content production.
        The mainstream approach so far is to develop individual algorithms for each domain and each task.
        In this paper, we explore the possibility of developing a unified foundation model
        to support \emph{open-ended domains and tasks} in an industrial recommender system, which may reduce the demand on downstream settings' data and can minimize the carbon footprint by avoiding training a separate model from scratch for every task.
        Deriving a unified foundation is challenging due to (i) the potentially unlimited set of downstream domains and tasks, and (ii) the real-world systems' emphasis on computational efficiency.
        We thus build our foundation upon M6, an existing large-scale industrial pretrained language model similar to GPT-3 and T5, and leverage M6's pretrained ability for sample-efficient downstream adaptation, by representing user behavior data as plain texts and converting the tasks to either language understanding or generation.
        To deal with a tight hardware budget, we propose an improved version of prompt tuning that outperforms fine-tuning with negligible 1\% task-specific parameters, and employ techniques such as late interaction, early exiting, parameter sharing, and pruning to further reduce the inference time and the model size.
        We demonstrate the foundation model's versatility on a wide range of tasks such as retrieval, ranking, zero-shot recommendation, explanation generation, personalized content creation, and conversational recommendation, and manage to deploy it on both cloud servers and mobile devices.
    \end{abstract}

    \maketitle

    \section{Introduction}\label{sec:intro}

Recommender systems are indispensable to many mobile applications and have become increasingly complex nowadays.
On one hand, an industrial recommender system is typically divided into \emph{a myriad of (sub)tasks}.
For example, there are tasks such as retrieval and ranking for deciding which contents to deliver to a user~\cite{youtubednn}, tasks such as keyword highlighting and explanation generation for polishing the way the contents are presented~\cite{explain-rec-survey}, and even tasks such as trend forecasting and AI-assisted content production for enriching the supply of contents~\cite{kang17visually}.
On the other hand, it is common for a single mobile application to have contents from \emph{diverse domains} due to business expansion.
For example, the recommendable contents in Taobao can be products, short videos or search queries, while Alipay also deals with multiple domains such as funds, mini-apps, articles, and search queries.

The mainstream paradigm to date remains developing independent algorithms for each subtask and building separate recommender systems for each domain.
This begs the question: can we establish a single foundation model that has open-ended ability, i.e., the ability to support a potentially unlimited set of downstream tasks and domains in recommender systems after minimal adaptation?
A foundation model for industrial recommender systems is attractive.
First, improvements in a unified foundation model can directly translate to all tasks and all domains.
It has been both empirically and theoretically demonstrated that an improvement in a single subtask alone may fail to bring any improvement in the system as a whole, e.g., in the case of bias reduction~\cite{fair-multi-recsys}, which can be circumvented if we optimize the foundation model to improve the whole system.
Second, a pretrained foundation model can reduce the need of task-specific data.
This benefits settings where data are scarce, for example, when expanding a business into a new domain or testing novel business ideas such as interactive recommendation or personalized content creation.
Third, a reusable foundation model could potentially reduce the carbon footprint.
The compute-intensive pretraining stage need only be executed once and adaptation to downstream tasks is computationally lightweight, unlike the existing paradigm where intensive computation is involved in every task and every domain.

However, establishing a foundation model for recommender systems is challenging.
It needs to support a wide range of domains and tasks with distinct characteristics, and faces strict restrictions on inference speed, memory consumption and storage requirement imposed by real-world systems.

In this paper, we extend our existing generative pretrained language model M6~\cite{m6-kdd,m6-arxiv,m6-10t} and present M6-Rec, a foundation model for recommender systems.
We convert all tasks in recommender systems into either language understanding or language generation.
Specifically, we collect anonymous user behavior data from recommender systems and store them as plain texts~\footnote{
    We focus primarily on texts.
    However, it is straightforward to support images by converting an image into a sequence of tokens, as in DALL-E~\cite{dall-e} and M6~\cite{m6-arxiv}.
}
such as ``A male user in Beijing, who clicked product X last night and product Y this noon, was recommended product Z and did not click it.''
We observe this text-based approach brings excellent zero/few-shot learning ability, thanks to the fact that M6 is pretrained on a comprehensive massive-scale corpus with well-designed objectives.

M6-Rec then makes several architectural changes to M6 for real-world deployment and for further reducing the carbon footprint.
First, we propose \emph{option tuning}, an improved variant of prompt tuning~\cite{prompt_tune}, in place of fine-tuning.
It adds negligible task-specific parameters and makes no modification to the pre-trained model's parameters, which alleviates catastrophic forgetting and makes it possible to deploy one single model for serving multiple tasks under a limited hardware budget via mixed-task inference~\cite{prompt_tune}.
Option tuning addresses prompt tuning's slow convergence, and is able to outperform fine-tuning when combined with adapter tuning~\cite{lora_adapter} despite tuning merely 1\% parameters.
Second, we implement a multi-segment variant of late interaction~\cite{colbert} when adapting M6-Rec to tasks that require low-latency real-time inference, where most of the heavy computation is pre-computed offline and cached.
Finally, to make M6-Rec deployable on edge devices such as mobile phones, we further employ techniques such as parameter sharing~\cite{albert}, pruning~\cite{pofa_bert}, quantization~\cite{q8bert}, and early-exiting~\cite{branchynet,msdnet} to reduce the model size and accelerate the inference speed.

In summary, our main contributions are:
\begin{itemize}
    \item We present M6-Rec, a foundation model that not only supports sample-efficient open-domain recommendation, but also unifies all subtasks in a recommender system ranging from content supply, delivery, to presentation.
    \item We improve prompt tuning to outperform fine-tuning while remaining parameter-efficient, and share experiences of making M6-Rec deployable in real-world systems via late interaction, parameter sharing, pruning, and early-exiting.
    \item M6-Rec supports zero/few-shot learning and performs well in abundant tasks and domains, ranging from classic tasks such as retrieval and ranking, to novel use cases such as personalized product design and conversational recommendation, and is deployed on both cloud servers and edge devices.
\end{itemize}

\begin{figure}
    \centering
    \includegraphics[width=0.5\textwidth]{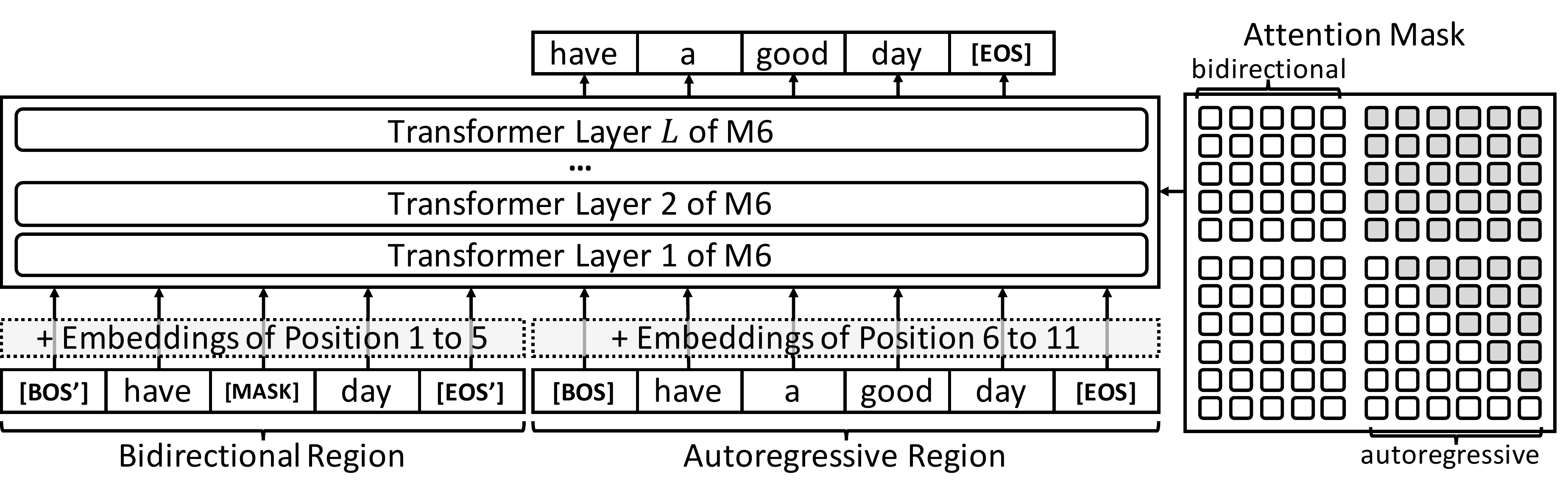}
    \caption{The text infilling objective used by the pretraining procedure of M6.
            {[MASK]} represents an undetermined number of unknown tokens. {[BOS]} and {[EOS]} mean the beginning and the end of a sentence, respectively.
        The autoregressive language modeling loss is imposed on the outputs of the autoregressive region, and not on the bidirectional region.
    }
    \label{fig:m6-pretrain}
\end{figure}

\begin{figure}
    \centering
    \includegraphics[width=0.5\textwidth]{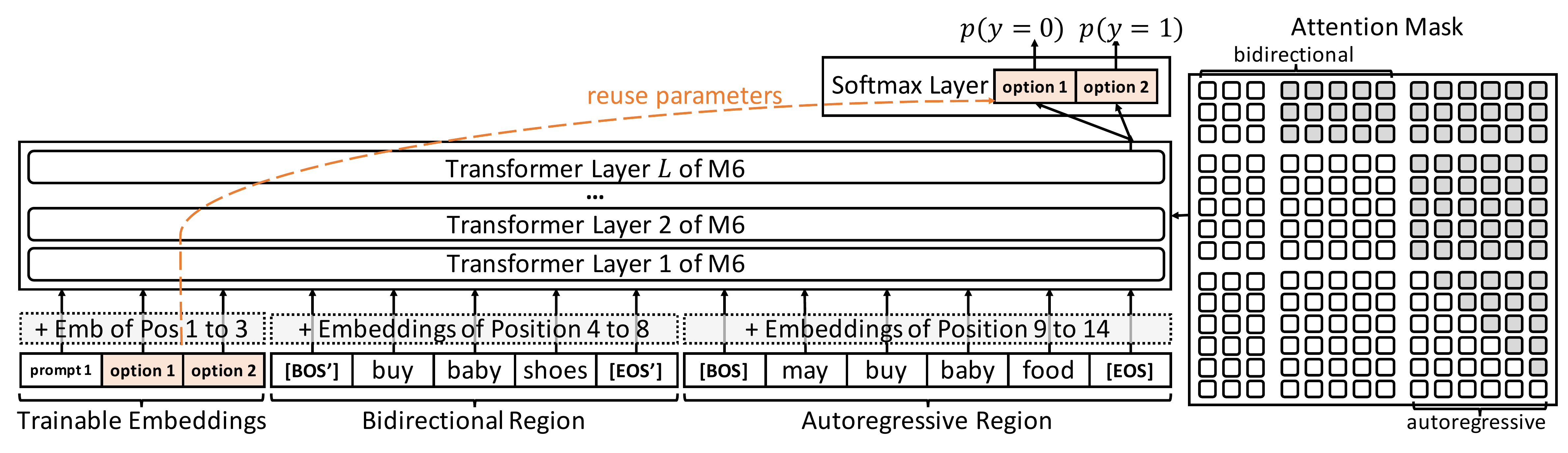}
    \caption{
        We propose option tuning, a variant of prompt tuning that enjoys better convergence.
        It reuses the last few soft prompts to serve as the parameters of the softmax classification layer, namely \emph{soft options}.
        The example in the figure has three soft prompts, two of which are soft options.
        It optimizes only the soft prompts and the special tokens such as {[EOS]}, while freezing the remaining parameters.
    }
    \label{fig:m6-option-tune}
\end{figure}

\begin{figure*}
    \centering
    \includegraphics[width=\textwidth]{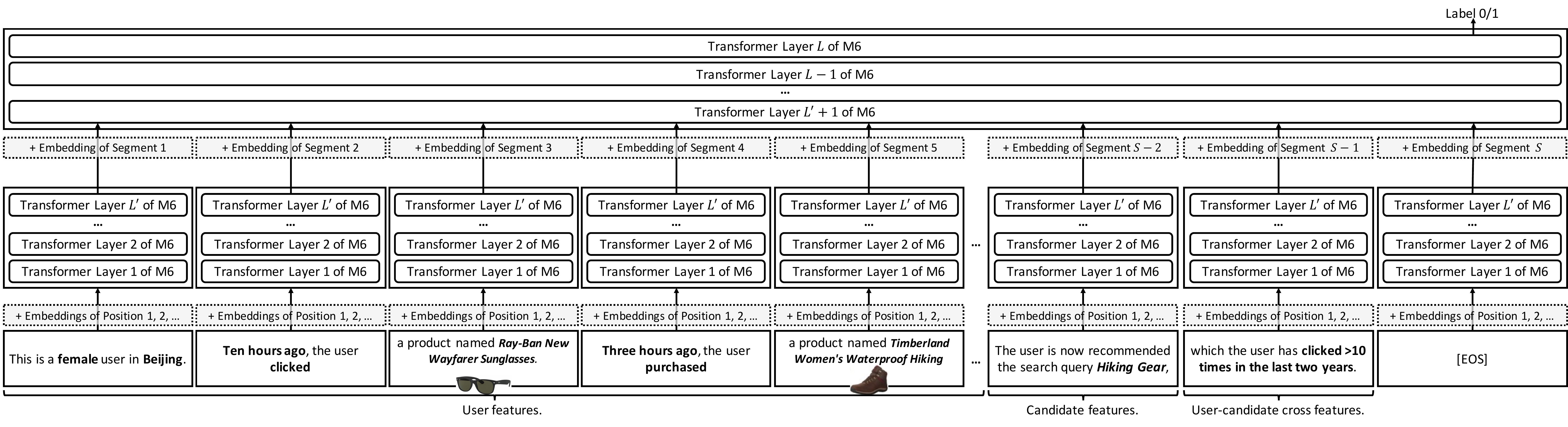}
    \caption{
        The figure showcases our implementation of late interaction for click-through rate (CTR) prediction.
        To support tasks that emphasize low-latency real-time inference, M6-Rec pre-computes and caches the first $L'$ layers' results, while computing the last $L-L'$ layers for performing late interaction when the request arrives.
        A user's features as a whole usually change frequently.
        We thus split a user's features into finer-grained segments that are more static, e.g., by representing each clicked item as an individual segment, so that we can dynamically incorporate the user's latest activities.
        The special token $[\mathrm{BOS}']$ before the user feature text and the two special tokens $[\mathrm{EOS}'][\mathrm{BOS}]$ before the candidate feature text are omitted for clarity.
    }
    \label{fig:late-iter}
\end{figure*}

\begin{figure}
    \centering
    \includegraphics[width=0.5\textwidth]{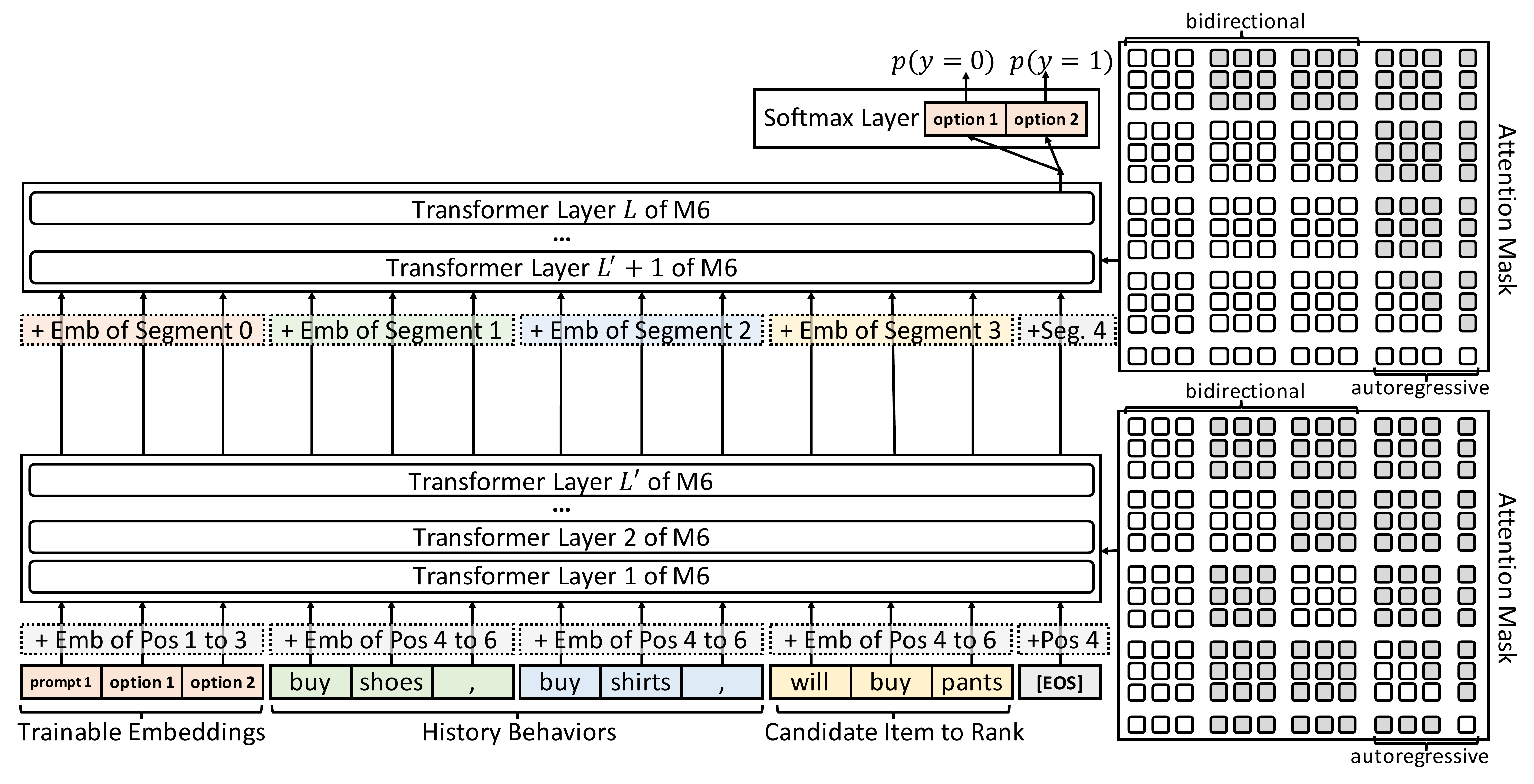}
    \caption{
        How to train M6-Rec using one single forward pass when option tuning and multi-segment late interaction are combined, rather than using separate forward passes for the multiple segments.
        This example has three soft prompts, two of which are soft options.
        The example user has two past behaviors, and we are predicting if the user will buy pants.
    }
    \label{fig:late-iter-traintime}
\end{figure}


    \section{Related Work}\label{sec:related}

In this section, we discuss the connection between M6-Rec and the existing works on recommender systems and pretrained models.

\subsection{Pretraining for Recommendation}\label{subsec:pretrain-rec}


The existing works~\cite{ptum,yuan2020parameter,yuan2021one} on pretraining for recommendation are mainly about transferable user modeling, e.g., transferring knowledge from large-scale scenarios to scenarios with lesser data~\cite{pretrain-rec}.
They have three major limitations:

First, the existing pretraining approaches for recommendation usually focus solely on behavior data and neglect non-behavior web corpora.
We believe that the knowledge mined from large-scale web corpora can complement behavior data and help reduce the reliance on behavior data.
For example, with the aid of web knowledge, a system may recommend turkeys when it is Thanksgiving even if we collect no click behaviors related with turkeys or Thanksgiving, i.e., in a zero-shot manner.
We thus build M6-Rec upon M6~\cite{m6-kdd,m6-arxiv}, an existing language model pretrained on web corpora.

Second, the existing approaches typically require that pretraining and downstream tasks involve the same set of items due to the use of item IDs' embeddings, which prevents them from getting wider adoption.
We thus decide to avoid item IDs in M6-Rec by using items' texts instead, and open up the possibility of open-domain recommendation where a target domain consists of unseen items.

Finally, the existing approaches deal only with scoring tasks, e.g., rating the compatibility between a user and an item, and ignore tasks that require generation, e.g., dialog-based recommendation where generating a text sentence is required.
Our M6-Rec tackles this limitation by building upon M6, a generative language model.


\subsection{Language Models as Foundations}\label{subsec:lang-foundation}

Language models such as BERT~\cite{bert}, GPT~\cite{gpt2,gpt3} and T5~\cite{2020t5} transfer knowledge from web corpora to downstream tasks and are thus deemed foundation models~\cite{foundation-models}.
Most of them are built upon the Transformer~\cite{transformer} because its training paradigm is parallelizable and costs less time.
There have been two recent trends in making language models more powerful foundations: (i) some researchers explore the scaling law and construct increasingly larger-scale models~\cite{megatron-lm,gshard,switchtrm,m6-10t,cpm2,pangu-alpha,ernie3}, and (ii) the others strive for the unification of multiple modalities~\cite{clip,dall-e,cogview,m6-arxiv,m6-ufc,vilbert,vl-bert}.

However, the research on combining language models and user behavior data remains less developed.
There are industrial works that fine-tune pretrained language models, for example based on BERT~\cite{twinbert} or ERNIE~\cite{ernie-rank,ernie-retrieval}, to produce representations for text contents and search queries.
Yet these works are limited in terms of supported domains and tasks, i.e., they target a single use case such as web searches and concern only scoring tasks, with no intention to support diverse domains or generation tasks.

The concurrent proof-of-concept work of Sileo et al.~\cite{zero-rec-lang} shows that a pretrained model such as GPT-2~\cite{gpt2} does contain knowledge useful for recommendation, by using a text prompt to guide the non-finetuned GPT-2 to rank movies.
We corroborate their findings and experiment with data from more domains.

WebGPT~\cite{webgpt} demonstrates that user behavior data are valuable for pretrained language models.
Specifically, WebGPT collects human behavior data in a simplified text-based web browsing environment, and use the behavior data for teaching GPT-3~\cite{gpt3} how to browse the web when answering questions.
Our work explores the opposite direction, i.e., we investigate if a pretrained language model can be a valuable foundation for behavior-related tasks.

\subsection{Efficient Language Foundations}\label{subsec:efficient-foundation}


\emph{Low-latency inference} is critical for a responsive user experience.
Early exiting~\cite{branchynet,msdnet} is a common technique for reducing the latency of deep models.
Previous works~\cite{bertpatience,deebert,dynabert} have adopted early exiting for BERT-like bidirectional language models, and we show in this work that it is also applicable to GPT-like autoregressive language models.
Late interaction~\cite{colbert} is another alternative in settings where caching intermediate results are possible, which segments Transformer into a deep stage that precomputes intermediate representations and a shallow stage for performing feature interaction.
Previous works~\cite{colbert,ernie-rank} consider the late interaction between two coarse-grained entities, and M6-Rec extends it to handle the interaction among multiple finer-grained segments.

\emph{Reducing the model size} keeps hardware costs down and is mandated for resource-limited edge devices.
Many strategies have been explored, e.g., parameter sharing~\cite{albert}, distillation~\cite{mobilebert,minilm,distillbert,tinybert}, pruning~\cite{lotterybert,compress_bert,pofa_bert}, and quantization~\cite{q8bert}.
Still, the existing tiny language models usually have over >10M parameters, while we estimate that it needs to be around 2M to avoid degrading the user experience when deploying a model to our users' mobile phones.

\emph{Parameter-efficient adaptation} introduces as few task-specific parameters as possible when adapting a pretrained model to downstream tasks.
Zero-parameter methods that use text prompts perform worse than full-model fine-tuning, but enjoy the merits of few- and zero-shot learning~\cite{gpt3}.
Other methods, such as adapters~\cite{lora_adapter}, prefix tuning~\cite{prefix_tune}, and prompt tuning~\cite{prompt_tune}, attempt to minimize the performance gap when reducing the number of task-specific parameters, and it is convenient to deploy one single model for serving multiple tasks simultaneously with these methods.
Concurrent to our work, He et al.~\cite{unify-eff-tune} unify adapters and prefix tuning, and discover that combining the two brings performance comparable to fine-tuning. 
Some researchers~\cite{transfer_prompt} have reported the slow convergence issue of prompt tuning.
We address the issue by adding soft options.
We also observe competitive performance after combining adapters and prompt tuning, even though prompt tuning cannot be precisely incorporated into He et al.'s unified framework.

    \vspace{-0.5cm}

\section{Method: M6-Rec}\label{sec:method}

Here we briefly describe M6-Rec's backbone M6 and give examples on how M6-Rec converts recommendation-related tasks into language understanding or generation.
We also propose option tuning and multi-segment late interaction for energy-efficient adaptation here, and share experiences on deploying M6-Rec to edge devices.

\subsection{M6: An Industrial-Strength Backbone}\label{subsec:prelim-m6}

M6~\cite{m6-kdd,m6-arxiv,m6-10t} is a series of visual-linguistic pretrained models.
We choose M6 as our backbone, because
(i) M6 supports both Chinese and English, (ii) M6 is a multi-modal model which aligns well with our plan to incorporate multi-modal features in the future, even though we primarily consider only texts in this paper for simplicity, and (iii) M6 has achieved widespread success in Alibaba Group's ecosystem when deployed into real-world businesses.

We use the base version of M6 that has around 300M parameters, unless otherwise stated.
M6-base consists of a single Transformer of $L=24$ layers, $H=16$ attention heads, and $d=1024$ hidden states.
It uses a sequence-to-sequence attention mask similar to UniLM~\cite{unilm}, where the source sentence falls within a region that uses the bidirectional mask and the target sentence is in a follow-up region that uses the unidirectional mask (see Figure~\ref{fig:m6-pretrain}).

Pretraining of M6 consists of (i) a text infilling objective~\cite{bart}, which benefits downstream scoring tasks by providing the ability to assess the plausibility of a text or an event, and (ii) an autoregressive language generation objective, which is critical for downstream text generation tasks.
M6 implements the infilling objective by masking small spans in a sentence and autoregressively generating the unmasked sentence given the masked one, whereas the language generation objective is implemented by masking the whole sentence.
We have omitted the visual modality of M6 here for clarity.

\subsection{Behavior Modeling as Language Modeling}\label{subsec:beh-as-lang}

We now illustrate how M6-Rec converts all \emph{downstream} tasks~\footnote{
    We have tried mixing the user behavior texts into M6's pretraining dataset and resuming \emph{pretraining} on the mixed dataset.
    However, this extra pretraining step brought only marginal gains in many cases and is thus not used for simplicity.
}
into language understanding or generation tasks by representing user behavior data as natural language plain texts for fine-tuning.

\paragraph{\bf Scoring tasks}
The most common tasks in a recommender system are about scoring the plausibility of an event, for example, click-through rate (CTR) prediction or conversion rate (CVR) prediction where the goal is to estimate the probability of a user clicking or purchasing an item.
M6-Rec expresses an example sample for CTR prediction as follows and sends it into M6's Transformer:
\begin{quote}
    $[\mathrm{BOS}']$
    December.
    Beijing, China.
    Cold weather.
    A male user in early twenties, searched ``winter stuff'' 23 minutes ago, clicked a product of category ``jacket'' named ``men's lightweight warm winter hooded jacket'' 19 minutes ago, clicked a product of category ``sweatshirt'' named ``men's plus size sweatshirt stretchy pullover hoodies'' 13 minutes ago, clicked \ldots \; $[\mathrm{EOS}']$
    $[\mathrm{BOS}]$
    The user is now recommended a product of category ``boots'' named ``waterproof hiking shoes mens outdoor''.
    The product has a high population-level CTR in the past 14 days, among the top 5\%.
    The user clicked the category 4 times in the last 2 years.
    $[\mathrm{EOS}]$
\end{quote}
The text between $[\mathrm{BOS}']$ and $[\mathrm{EOS}']$ describes the user-side features, which corresponds to
the first region of M6's input, i.e., the bidirectional region in Figure~\ref{fig:m6-pretrain}.
The text between $[\mathrm{BOS}]$ and $[\mathrm{EOS}]$ describes the features related with the candidate item as well as the features related with both the user and the candidate item, which corresponds to the second region of M6's input, i.e., the autoregressive region in Figure~\ref{fig:m6-pretrain}.
M6-Rec then uses the output vector of M6's Transformer at $[\mathrm{EOS}]$'s position for summarizing the sample, sends the vector into a linear softmax classifier (which has two classes in the case of CTR prediction), and minimizes the cross-entropy loss.

\paragraph{\bf Generation tasks}
Content generation has become an important topic in modern recommender systems.
M6-Rec uses the following plain text format to support both personalized product design~\cite{kang17visually} and explainable recommendation~\cite{explain-rec-survey}:
\begin{quote}
    $[\mathrm{BOS}']$ \ldots $[\mathrm{EOS}']$
    $[\mathrm{BOS}]$
    The user now purchases a product of category ``\ldots'' named ``\ldots''. \;
    Product details: \ldots \;
    The user likes it because \ldots \;
    $[\mathrm{EOS}]$
\end{quote}
The omitted text between $[\mathrm{BOS}']$ and $[\mathrm{EOS}']$ describes a user's attributes and past behaviors.
The omitted text after ``Product details:'' is the detailed description of the product provided the seller.
The product description is included here to provide basic facts for drafting an explanation text when making a recommendation decision.
The omitted text after ``because'' is a brief sentence summarizing the user's opinion on the product, mined from the user's review using existing tools such as aspect-based sentiment analysis.
The text after ``because'' serves as the ground-truth for explanation generation.
Given a few samples of this format, M6-Rec tunes M6's Transformer using the same autoregressive language modeling loss as that of the pretraining stage shown in ~\ref{fig:m6-pretrain}.

\paragraph{Generation task: explanation generation}
We provide the input text prior to and including the word ``because'' during inference, and let the model generates the follow-up text.
We can then use the generated text as the explanation for the recommendation.

\paragraph{Generation task: personalized product design}
This task is equivalent to filling in the two blanks in the text ``$[\mathrm{BOS}']$ \ldots $[\mathrm{EOS}']$ $[\mathrm{BOS}]$ The user now purchases a product of category \rule{0.5cm}{0.15mm} named \rule{0.5cm}{0.15mm}''.
The model thus needs to predict the category name and the product's title based on the user's attributes and past behaviors, which helps us identify the keywords that the user may potentially be interested in.
Explicitly providing the category can force M6-Rec to design a product of that category.
We can further send the title into a text-to-image synthesis pipeline such as M6-UFC~\cite{m6-ufc}.

\paragraph{Generation task: personalized search query generation}
M6-Rec implements it as filling in the blank of ``$[\mathrm{BOS}']$ \ldots $[\mathrm{EOS}']$ $[\mathrm{BOS}]$ The user now searches the query \rule{0.5cm}{0.15mm}'' based on a user's past behaviors.

\paragraph{Generation task: conversational recommendation}
M6-Rec supports this task by marking the speaker of each sentence:
\begin{quote}
    $[\mathrm{BOS}']$ \ldots $[\mathrm{EOS}']$
    $[\mathrm{BOS}]$
    USER: Hi!
    SYSTEM: What kind of movie do you like?
    USER: I like horror movies.
    SYSTEM: How about \emph{The Shining (1980)}?
    \ldots
    $[\mathrm{EOS}]$
\end{quote}
We may put information about the user along with some basic facts between $[\mathrm{BOS}']$ and $[\mathrm{EOS}']$ if necessary.

\paragraph{\bf Zero-shot scoring tasks}
The format we described previously for scoring tasks works well if we have a few samples for training.
However, it does not support zero-shot learning due to the need of training a classification layer.
Fortunately, language models can estimate the likelihood of an event in a zero-shot manner, as long as the event is described in natural language.
For example, to estimate whether a user who clicks hiking shoes prefers trekking poles or yoga knee pads,
we can construct the following two sentences:
\begin{quote}
    $[\mathrm{BOS}']$ A user clicks hiking shoes $[\mathrm{EOS}']$
    $[\mathrm{BOS}]$ also clicks trekking poles $[\mathrm{EOS}]$

    $[\mathrm{BOS}']$ A user clicks hiking shoes $[\mathrm{EOS}']$
    $[\mathrm{BOS}]$ also clicks yoga knee pads $[\mathrm{EOS}]$
\end{quote}
M6-Rec sends each sentence into M6's Transformer and obtains the probability of each token.
Let $p_1, p_2, p_1', p_2', p_3'$ be the output probabilities corresponding to token ``trekking'', ``poles'', ``yoga'', ``knee'', ``pads'', respectively.
M6-Rec compares the plausibility of the two events by comparing $\frac{\log p_1 + \log p_2}{2}$ and $\frac{\log p_1' + \log p_2' + \log p_3'}{3}$.
Here we have normalized the log likelihood by the number of tokens.

M6-Rec supports other zero-shot tasks, such as building tag-based user profiles in a zero-shot manner, in a similar vein.

\paragraph{\bf Retrieval tasks}
Industrial recommender systems include a task called retrieval, also known as matching or candidate generation, which is executed prior to ranking tasks such as CTR and CVR prediction~\cite{youtubednn}.
Retrieval is about retrieving a small subset of candidate items from a pool of millions or even billions of items, usually via k-nearest neighbor (kNN) search after projecting the users and items into a vector space.
M6-Rec feeds the information of a user into M6's Transformer using the following format:
\begin{quote}
    $[\mathrm{BOS}']$
    December.
    Beijing, China.
    Cold weather.
    A male user in early twenties, searched ``winter stuff'' 23 minutes ago, clicked a product of category ``jacket'' named ``men's lightweight warm winter hooded jacket'' 19 minutes ago, \ldots \; $[\mathrm{EOS}']$
    $[\mathrm{BOS}]$
    $[\mathrm{EOS}]$
\end{quote}
The output corresponding to $[\mathrm{EOS}']$ is linearly projected into a $128$-dimensional vector and $l_2$-normalized to form the user's vector representation $\mathbf{x}$.
An item's text is fed into the model as follows:
\begin{quote}
    $[\mathrm{BOS}']$
    $[\mathrm{EOS}']$
    $[\mathrm{BOS}]$
    A product of category ``boots'' named ``waterproof hiking shoes mens outdoor''.
    High CTR among the top 5\%.
    Product details: \ldots
    $[\mathrm{EOS}]$
\end{quote}
Again, the output at $[\mathrm{EOS}]$ is projected into a $128$-dimensional vector
and $l_2$-normalized to serve as the item's vector representation $\mathbf{y}$.
We use two different learnable matrices for linearly projecting user vectors and item vectors, respectively.
And $l_2$-normalization enables more accurate and efficient kNN search.
M6-Rec uses contrastive learning to tune the model, by minimizing
\begin{align*}
    \sum_{\langle \mathbf{x}, \mathbf{y} \rangle} -\log
    \frac{
        \exp \left( \mathbf{x}^\top \mathbf{y} / \tau \right)
    }{
        \exp \left( \mathbf{x}^\top \mathbf{y} / \tau \right)
        + \sum_{\mathbf{y}' \in \mathcal{Y}'} \exp \left( \mathbf{x}^\top \mathbf{y}' / \tau \right)
    },
\end{align*}
where $\langle \mathbf{x}, \mathbf{y} \rangle$ is a positive pair, for example if user $\mathbf{x}$ has clicked item $\mathbf{y}$, while $\mathcal{Y}'$ is a set of randomly sampled items.
Following the literature~\cite{clrec}, we use the items in the same mini-batches to form $\mathcal{Y}'$ and set the temperature hyperparameter $\tau$ to $0.07$.

\subsection{Late Interaction for Low-Latency Inference}\label{subsec:late-inter}

Some components in recommender systems have a strict restraint on inference latency in order to meet users' real-time needs.
Techniques such as distillation~\cite{mobilebert,minilm,distillbert,tinybert} and early exiting~\cite{bertpatience,deebert} can avoid high latency, but cannot fully leverage a giant foundation model's capacity since they reduce or use only a subset of parameters.
We thus design \emph{multi-segment late interaction}, an extended variant of the original late interaction~\cite{colbert}, to cope with the issue.

We illustrate the design of multi-segment late interaction in Figure~\ref{fig:late-iter} using CTR prediction as an example.
The core idea is to pre-compute and cache the computation results of the Transformer's first $L'$ layers, and run the Transformer's last $L-L'$ layers in real time when a user request arrives.
We use $L-L'\le 3$ for low latency.

M6-Rec with late interaction does not run the first $L'$ layers directly on the whole request when processing a request such as ``Male. Clicked X. Clicked Y. Will click Z?''.
Rather, it chunks the request into several segments, i.e., ``Male.'', ``Clicked X.'', ''Clicked Y.'', and ``Will click Z?''.
It processes the segments individually, and caches the segments' results for future reuse.
Segmentation brings high reusability.
For instance, when processing another request such as ``Female. Clicked Y. Will click Z?'', the already cached results of segment ``Clicked Y.'' and segment ``Will click Z?'' are reused.

The first $L'$ layers do not model the interaction among the segments of a request.
We hence concatenate these segments' results precomputed by the first $L'$ layers to form a single sequence, and run the last $L-L'$ layers of M6's Transformer on the said sequence to take the interaction into account.
Note that the first token in every segment is always considered as if it is at position 1, because the segments are processed separately by the first $L'$ layers.
As a result, we need to superimpose a set of learnable embeddings on the first $L'$ layers' outputs to let the last $L-L'$ layers know which are from the first segment and which are from the second segment, etc., which is similar to how position embeddings work.

\subsection{Option Tuning for Efficient Adaptation}\label{subsec:efficient-adapt}

We propose option tuning, which improves upon prompt tuning~\cite{prompt_tune} by addressing the slow convergence issue reported by some researchers~\cite{transfer_prompt}.
We then propose option-adapter tuning, which augments option tuning with adapters~\cite{lora_adapter} and outperforms fine-tuning despite tuning orders of magnitudes less parameters.

Prompt tuning prepends the input sequence with a predetermined number of trainable embeddings that serve as ``soft prompts'' in place of the discrete text prompts used by GPT-3, and tunes the soft prompts for downstream adaptation while freezing the pretrained model's parameters.
Our variant of prompt tuning reuses the last $C$ soft prompts (depicted as ``option embeddings'' in Figure~\ref{fig:m6-option-tune}) to serve as the softmax classification layer's parameters, where $C$ is the number of classes.
Our approach empirically converges better than using a separate learnable linear classification layer.
We name our approach \emph{option tuning}, since it is in spirit providing the answer options in the soft prompts describing the downstream task.

We then combine option tuning and adapter tuning.
Specifically, we add a low-rank adapter to the feedforward network (FFN) of every Transformer layer.
That is, the feedforward network at the $l$-th layer $\mathrm{FFN}^{(l)}(\cdot)$ is replaced by $\overline{\mathrm{FFN}}^{(l)}(\cdot)$ as follows:
\begin{align*}
    \overline{\mathrm{FFN}}^{(l)}(\mathbf{Z}) =
    \mathrm{FFN}^{(l)}(\mathbf{Z}) +
    \lambda^{(l)}
    \cdot \left[
    \sigma (\mathbf{Z} \; \mathbf{W}_{1}^{(l)} + \mathbf{b}_{1}^{(l)} ) \;
    \mathbf{W}_{2}^{(l)} + \mathbf{b}_{2}^{(l)}
    \right],
\end{align*}
where
$\mathbf{Z}\in \mathbb{R}^{B\times d}$ is the input of the FFN, $B$ is the batch size, and
$
\lambda^{(l)} \in \mathbb{R},
\mathbf{W}_{1}^{(l)} \in \mathbb{R}^{d\times r},
\mathbf{b}_{1}^{(l)} \in \mathbb{R}^{1\times r},
\mathbf{W}_{2}^{(l)} \in \mathbb{R}^{r\times d},
\mathbf{b}_{2}^{(l)} \in \mathbb{R}^{1\times d}
$.
are the trainable parameters of the adapter.
Typically $r \ll d$.
We name this hybrid approach \emph{option-adapter tuning}.

\subsection{Model Compression for Mobile Devices}\label{subsec:model-compress}

One of the frontiers of modern recommender system is edge computing, i.e., deploying algorithms on IoT devices or users' mobile phones to ensure responsiveness and protect user privacy.
However, edge devices have varying and usually limited resources.

To reduce the model size, we use MiniLM's relation-based approach~\cite{minilm} to distill the 300M M6-base into a 10M tiny model which is named M6-Edge.
Distillation is performed during the pretraining stage rather than the tuning stage.
M6-Edge uses the same input-output schema and pretraining tasks as M6-base.
M6-Edge has $L=24$ layers, $H=16$ attention heads, and $768$ hidden states.
It follows ALBERT~\cite{albert} in that (i) the $24$ layers share the same set of parameters and (ii) the token embeddings are linearly projected to a space of $768$ dimensions from $128$-dimensional embeddings.

We then perform a post-distillation pretraining step to further reduce the size and inference time of M6-Edge, using pruning and early exiting.
We use gradual magnitude pruning~\cite{gradualmagprune} for unstructured weight pruning.
For every embedding table and every linear layer, we gradually prune its weights of the lowest magnitude until reaching an 80\% sparsity ratio, while the training loss is continuously optimized in the process.
This reduces M6-Edge's size from 10M to 2M.
When early exiting at layer $k$, the outputs of the $k$-th layer instead of the final layer are used for training and inference.
Let $\mathcal{L}_k$ be the loss computed when early exiting at layer $k$.
In this post-distillation pretraining step, we follow the literature~\cite{bertpatience} and optimize the accumulated loss $\sum_{k=1}^{L} \frac{2 k}{k (k+1)} \mathcal{L}_k$ instead of just $\mathcal{L}_{L}$.
The accumulated loss is also used for downstream tasks.

Some low-end devices can only afford one or two layers while other high-end ones may use more.
Early exiting-based inference can use as many layers as possible depending on the hardware.
We also perform 8-bit quantization~\cite{q8bert} when deploying M6-Edge.

    \begin{table*}
    \centering
    \caption{
        Performance on explainable recommendation.
        R1-P, R1-R, R1-F, R2-P, R2-R, and R2-F denote the Precision, Recall, and F1 scores of ROUGE-1 and ROUGE-2, and $\uparrow$ indicates higher scores are better while $\downarrow$ indicates the opposite.
    }
    \label{tab:explain-rec}
    \begin{tabular}{@{}crrrrrrrrrrrr@{}}
        \toprule
        & \multicolumn{3}{c}{{\bf Explainability}}
        & \multicolumn{9}{c}{{\bf Text Quality}}
        \\
        \cmidrule(l){2-4}
        \cmidrule(l){5-13}
        {\bf Method}
        & {\bf FMR$\uparrow$} & {\bf FCR$\uparrow$} & {\bf DIV$\downarrow$}
        & {\bf USR$\uparrow$} & {\bf BLEU-1$\uparrow$} & {\bf BLEU-4$\uparrow$}
        & {\bf R1-P$\uparrow$} & {\bf R1-R$\uparrow$} & {\bf R1-F$\uparrow$}
        & {\bf R2-P$\uparrow$} & {\bf R2-R$\uparrow$} & {\bf R2-F$\uparrow$}
        \\
        \midrule
        Transformer~\cite{transformer} & 0.10 & 0.01 & 3.26 & 0.00 & 9.71 & 0.59 & 19.68 & 11.94 & 14.11 & 2.10 & 1.39 & 1.55
        \\
        NRT~\cite{nrt} & 0.12 & 0.07 & 2.93 & 0.17 & 12.93 & 0.96 & 21.03 & 13.57 & 15.56 & 2.71 & 1.84 & 2.05
        \\
        Att2Seq~\cite{att2seq} & 0.12 & 0.20 & 2.74 & 0.33 & 12.56 & 0.95 & 20.79 & 13.31 & 15.35 & 2.62 & 1.78 & 1.99
        \\
        PETER~\cite{ACL21-PETER} & 0.12 & 0.21 & 1.75 & 0.29 & 12.77 & 1.17 & 19.81 & 13.80 & 15.23 & 2.80 & 2.08 & 2.20
        \\
        ACMLM~\cite{ACMLM} & 0.10 & 0.31 & 2.07 & 0.96 & 9.52 & 0.22 & 11.65 & 10.39 & 9.69 & 0.71 & 0.81 & 0.64
        \\
        NETE~\cite{CIKM20-NETE} & 0.71 & 0.19 & 1.93 & 0.57 & 18.76 & 2.46 & 33.87 & 21.43 & 24.81 & 7.58 & 4.77 & 5.46
        \\
        PETER+~\cite{ACL21-PETER} & 0.77 & 0.31 & 1.20 & 0.46 & 19.75 & 3.06 & 34,71 & 23.99 & 26.35 & 9.04 & 6.23 & 6.71
        \\
        M6-Rec & {\bf 0.98} & {\bf 0.44} & {\bf 0.89} & {\bf 0.89} & {\bf 20.38} & {\bf 3.59}
        & {\bf 43.77} & {\bf 33.02} & {\bf 34.16} & {\bf 17.91} & {\bf 13.73} & {\bf 13.78}
        \\
        \bottomrule
    \end{tabular}
\end{table*}

\begin{table}
    \centering
    \caption{
        Performance on click-through rate (CTR) prediction, measured in terms of AUC.
    }
    \label{tab:ctr}
    \begin{tabular}{@{}ccrr@{}}
        \toprule
        & & \multicolumn{2}{c}{{\bf Datasets}}
        \\
        \cmidrule(l){3-4}
        {\bf Method} & {\bf Method Type} & {\bf AlipayQuery$\uparrow$} & {\bf TaoProduct$\uparrow$}
        \\
        \midrule
        DIN & ID embeddings & 0.7332 & 0.7611
        \\
        M6-Rec & Text semantics & {\bf 0.7508} & {\bf 0.7995}
        \\
        \bottomrule
    \end{tabular}
\end{table}

\begin{table}
    \centering
    \caption{
        Performance on the kNN retrieval task in terms of HitRate@100, conducted on the AlipayMiniApp dataset.
        We report the performance on all the test cases as well as the performance on the test cases that involve unseen items, e.g., cases where users have searched new queries or cases where the candidate mini-apps do not appeared in the training set.
    }
    \label{tab:retrieval}
    \begin{tabular}{@{}cccr@{}}
        \toprule
        & & \multicolumn{2}{c}{{\bf Test Sets}}
        \\
        \cmidrule(l){3-4}
        {\bf Method} & {\bf Method Type} & {\bf All Items$\uparrow$} & {\bf Unseen Items$\uparrow$}
        \\
        \midrule
        YouTubeDNN & ID embeddings & 54.4\% & \emph{fail}
        \\
        TwinBERT & Text semantics & 69.6\% & 49.6\%
        \\
        M6-Rec & Text semantics & {\bf 74.1\%} & {\bf 57.0\%}
        \\
        \bottomrule
    \end{tabular}
\end{table}

\begin{table}
    \centering
    \caption{
        Performance on personalized content creation on dataset AlipayQuery, where the goal is to generate search queries in real time for recommending to a user.
    }
    \label{tab:query-gen}
    \begin{tabular}{@{}crrrrr@{}}
        \toprule
        & \multicolumn{5}{c}{{\bf Metrics}}
        \\
        \cmidrule(l){2-6}
        {\bf Method}
        & {\bf Dist-2$\uparrow$} & {\bf Dist-3$\uparrow$} & {\bf Dist-4$\uparrow$}
        & {\bf BLEU-2$\uparrow$} & {\bf BLEU-3$\uparrow$}
        \\
        \midrule
        KOBE & 0.086 & 0.106 & 0.129 & 0.262 & 0.222
        \\
        M6-Rec & {\bf 0.093} & {\bf 0.115} & {\bf 0.148} & {\bf 0.278} & {\bf 0.244}
        \\
        \bottomrule
    \end{tabular}
\end{table}

\begin{table}
    \centering
    \caption{
        Performance on conversational recommendation.
    }
    \label{tab:dialog-rec}
    \begin{tabular}{@{}crrrrr@{}}
        \toprule
        & \multicolumn{5}{c}{{\bf Metrics}}
        \\
        \cmidrule(l){2-6}
        {\bf Method}
        & {\bf PPL$\downarrow$} & {\bf BLEU-2$\uparrow$} & {\bf BLEU-3$\uparrow$}
        & {\bf Dist-3$\uparrow$} & {\bf Dist-4$\uparrow$}
        \\
        \midrule
        Transformer & 20.44 & 0.026 & 0.014 & 0.27 & 0.39
        \\
        KBRD~\cite{kbrd} & 17.90 & 0.060 & 0.024 & 0.30 & 0.45
        \\
        KGSF~\cite{kgsf} & 10.73 & 0.033 & 0.022 & 0.40 & 0.46
        \\
        M6-Rec & {\bf 10.25} & {\bf 0.122} & 0.021 & {\bf 0.46} & {\bf 0.64}
        \\
        \bottomrule
    \end{tabular}
\end{table}

\begin{table}
    \centering
    \caption{
        Effects of multi-segment late interaction for CTR prediction, conducted on TaoProduct.
    }
    \label{tab:late-inter-speed}
    \begin{tabular}{@{}ccc@{}}
        \toprule
        {\bf Method} & {\bf AUC$\uparrow$} & {\bf Latency (ms)$\downarrow$}
        \\
        \midrule
        M6-Rec ($L=24$) & 0.7995 & 57
        \\
        \midrule
        M6-Rec, distilled ($L=3$) & 0.7566 & 16
        \\
        M6-Rec, late-inter ($L=24, L-L'=1$) & 0.7299 & 10
        \\
        M6-Rec, late-inter ($L=24, L-L'=3$) & {\bf 0.7731} & 16
        \\
        \bottomrule
    \end{tabular}
\end{table}

\begin{table}
    \centering
    \caption{
        Performance of our aggressively distilled tiny foundation model for edge deployment, i.e., M6-Edge.
    }
    \label{tab:edge-model-perf}
    \begin{tabular}{@{}crrrr@{}}
        \toprule
        & & \multicolumn{3}{c}{{\bf Tasks}}
        \\
        \cmidrule(l){3-5}
        {\bf Method} & {\bf \#Params}
        & {\bf TNEWS$\uparrow$} & {\bf IFLYTEK$\uparrow$} & {\bf CSL$\uparrow$}
        \\
        \midrule
        M6-base & 327M & 0.598 & 0.631 & 0.852
        \\
        \midrule
        ALBERT-zh-base & 12M & 0.550 & 0.564 & 0.785
        \\
        M6-Edge & 10M & {\bf 0.552} & {\bf 0.586} & {\bf 0.831}
        \\
        \midrule
        ALBERT-zh-tiny & 4M & 0.534 & 0.488 & 0.750
        \\
        M6-Edge, Pruned & 2M & {\bf 0.537} & {\bf 0.559} & {\bf 0.798}
        \\
        \bottomrule
    \end{tabular}
\end{table}

\begin{table}
    \centering
    \caption{
        Ablation study of our tuning methods, i.e.\ option tuning and option-adapter tuning, on the CLUE benchmark using the distilled 10M M6-Edge.
        Adding soft options to prompt tuning leads to option tuning, and option-adapter tuning is equal to prompt tuning with soft options and adapters.
    }
    \label{tab:prompt-tune}
    \begin{tabular}{@{}crrrr@{}}
        \toprule
        & & \multicolumn{3}{c}{{\bf Tasks}}
        \\
        \cmidrule(l){3-5}
        {\bf Tuning Method} & {\bf \#Params}
        & {\bf TNEWS$\uparrow$} & {\bf IFLYTEK$\uparrow$} & {\bf CSL$\uparrow$}
        \\
        \midrule
        Fine-Tuning & 100\% & 0.544 & 0.575 & 0.829
        \vspace{1mm}
        \\
        Adapter Tuning & 1\% & 0.542 & 0.574 & 0.825
        \vspace{1mm}
        \\
        Prompt Tuning & 1\% & 0.534 & 0.531 & 0.760
        \vspace{1mm}
        \\
        \begin{tabular}{@{}l@{}}
            Prompt Tuning    \\ {\small +Soft Options}
        \end{tabular}
        & 1\% & 0.544 & 0.565 & 0.813
        \vspace{1mm}
        \\
        \begin{tabular}{@{}l@{}}
            Prompt Tuning    \\ {\small +Soft Options} \\ {\small +FFN Adapters}
        \end{tabular}
        & 1\% & {\bf 0.552} & {\bf 0.586} & {\bf 0.831}
        \\
        \bottomrule
    \end{tabular}
\end{table}

\begin{figure}
    \centering
    \includegraphics[width=0.3\textwidth]{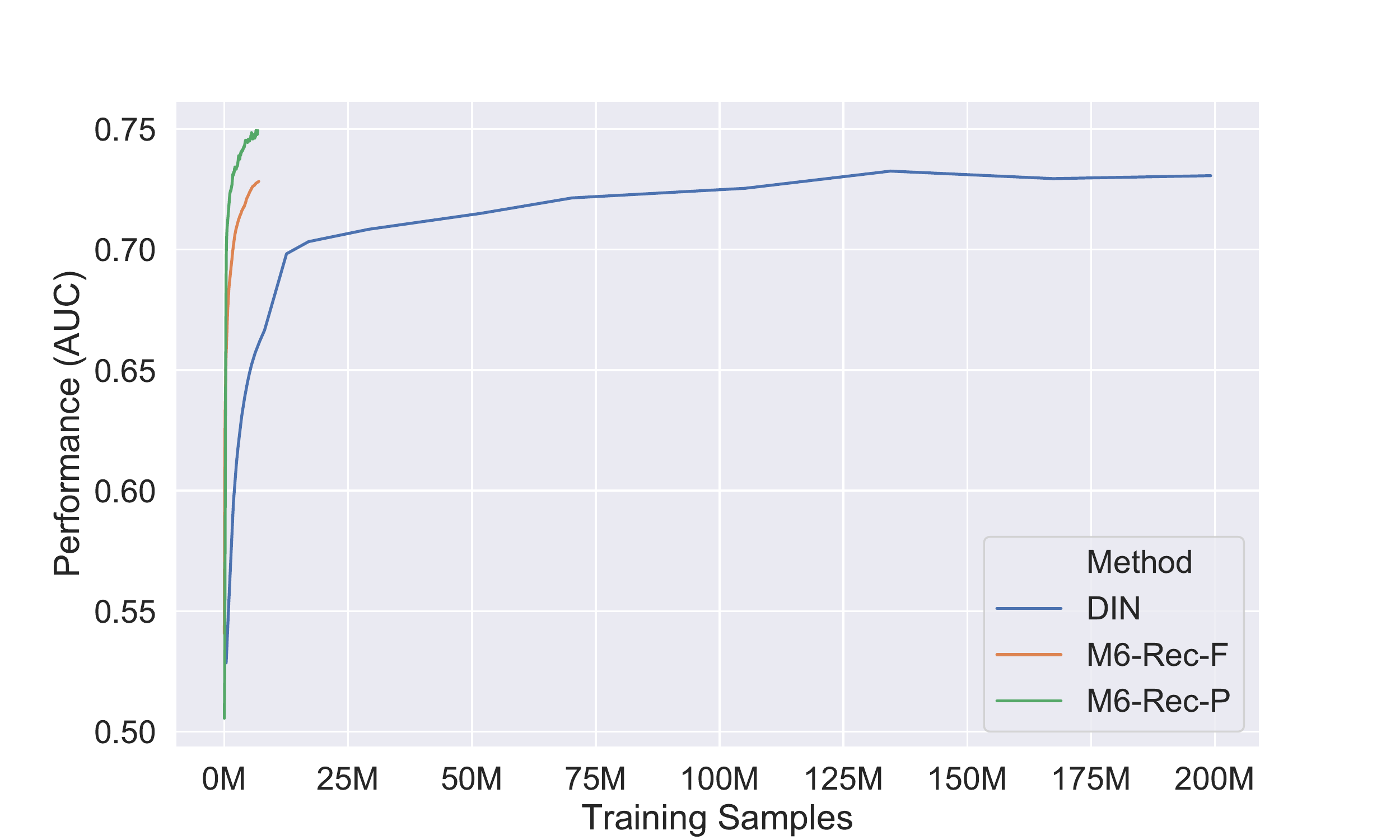}
    \caption{
        Performance on CTR prediction on AlipayQuery when the amount of data varies.
        M6-Rec-F and M6-Rec-P tune the model using fine-tuning and the variant of prompt tuning proposed by us (i.e.\ option-adapter tuning), respectively.
    }
    \label{fig:ctr-data-scale}
\end{figure}

\begin{figure}
    \centering
    \includegraphics[width=0.5\textwidth]{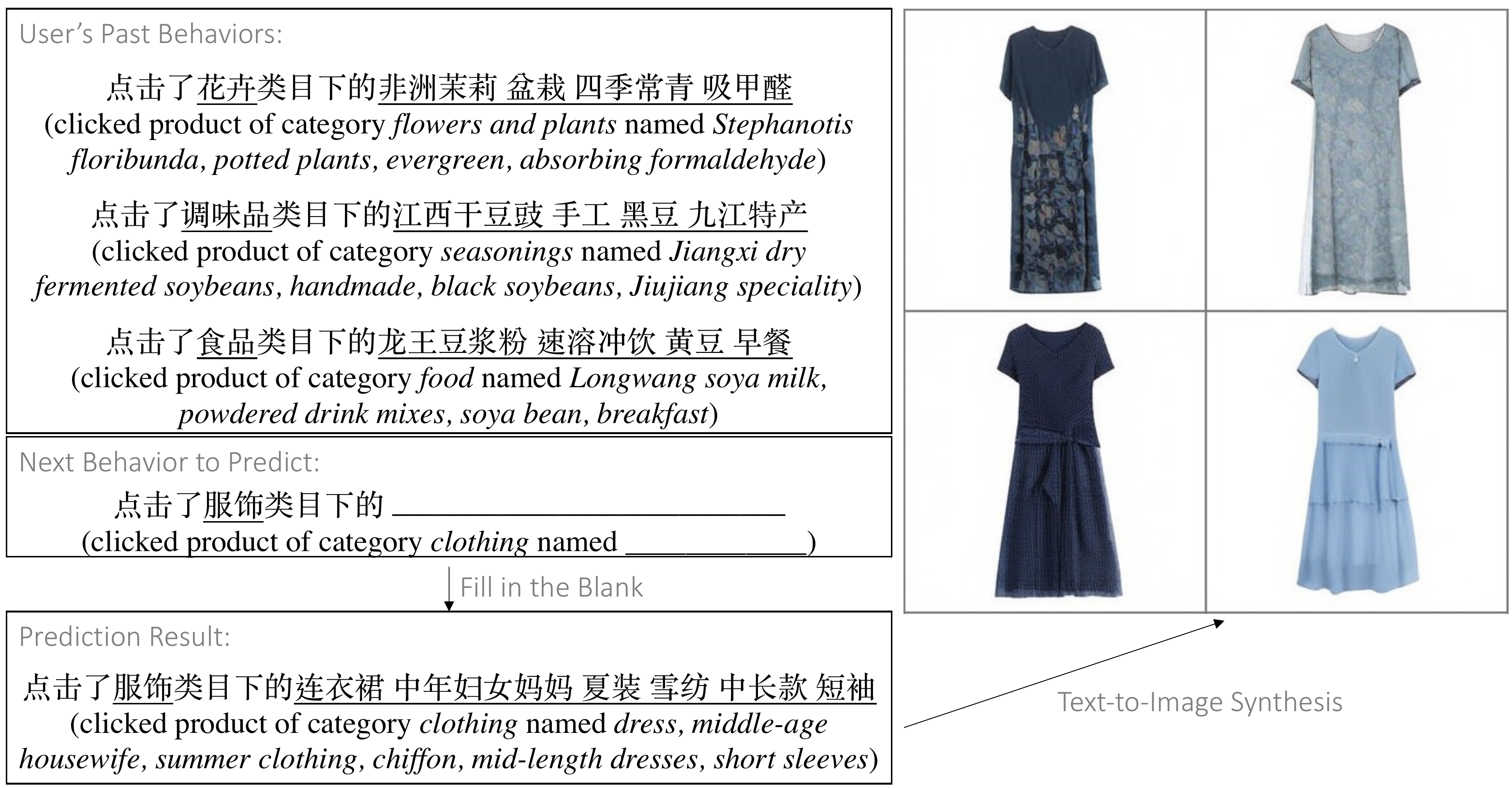}
    \caption{
        Personalized product design for Taobao.
        M6-Rec generates the ideal product title based on a user's past behaviors.
        The text-to-image synthesis pipeline of M6-UFC~\cite{m6-ufc} is then used.
        M6-Rec predicts that this example user is a middle-aged housewife based on the user's past behaviors.
    }
    \label{fig:tao-product-gen}
\end{figure}

\begin{figure}
    \centering
    \subfigure[On dataset AlipayQuery.]{
        \centering
        \includegraphics[width=0.28\textwidth]{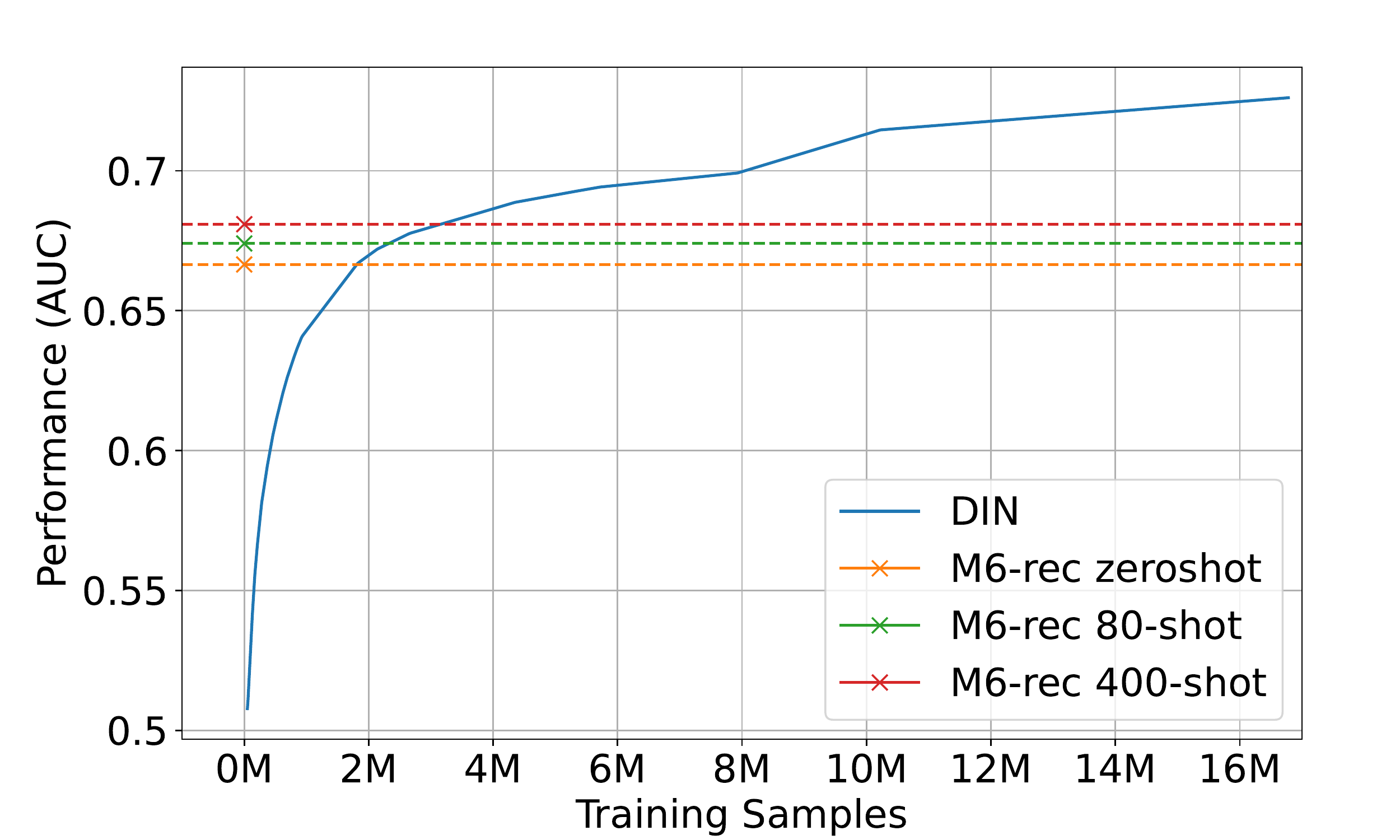}
    }
    \subfigure[On dataset Amazon-Movie~\cite{amazon-data-2019}.]{
        \centering
        \includegraphics[width=0.28\textwidth]{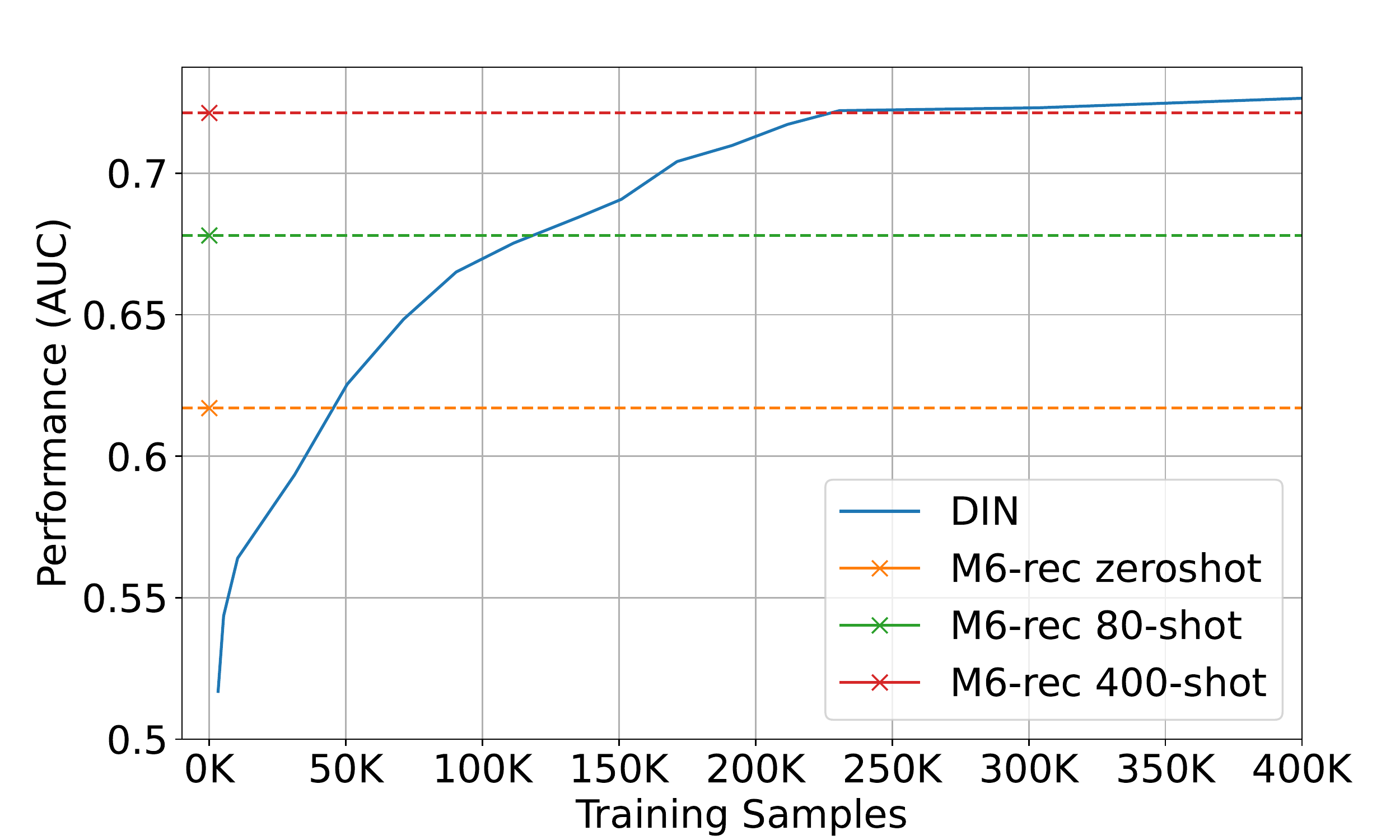}
    }
    \subfigure[On dataset Amazon-Cloth~\cite{amazon-data-2019}.]{
        \centering
        \includegraphics[width=0.28\textwidth]{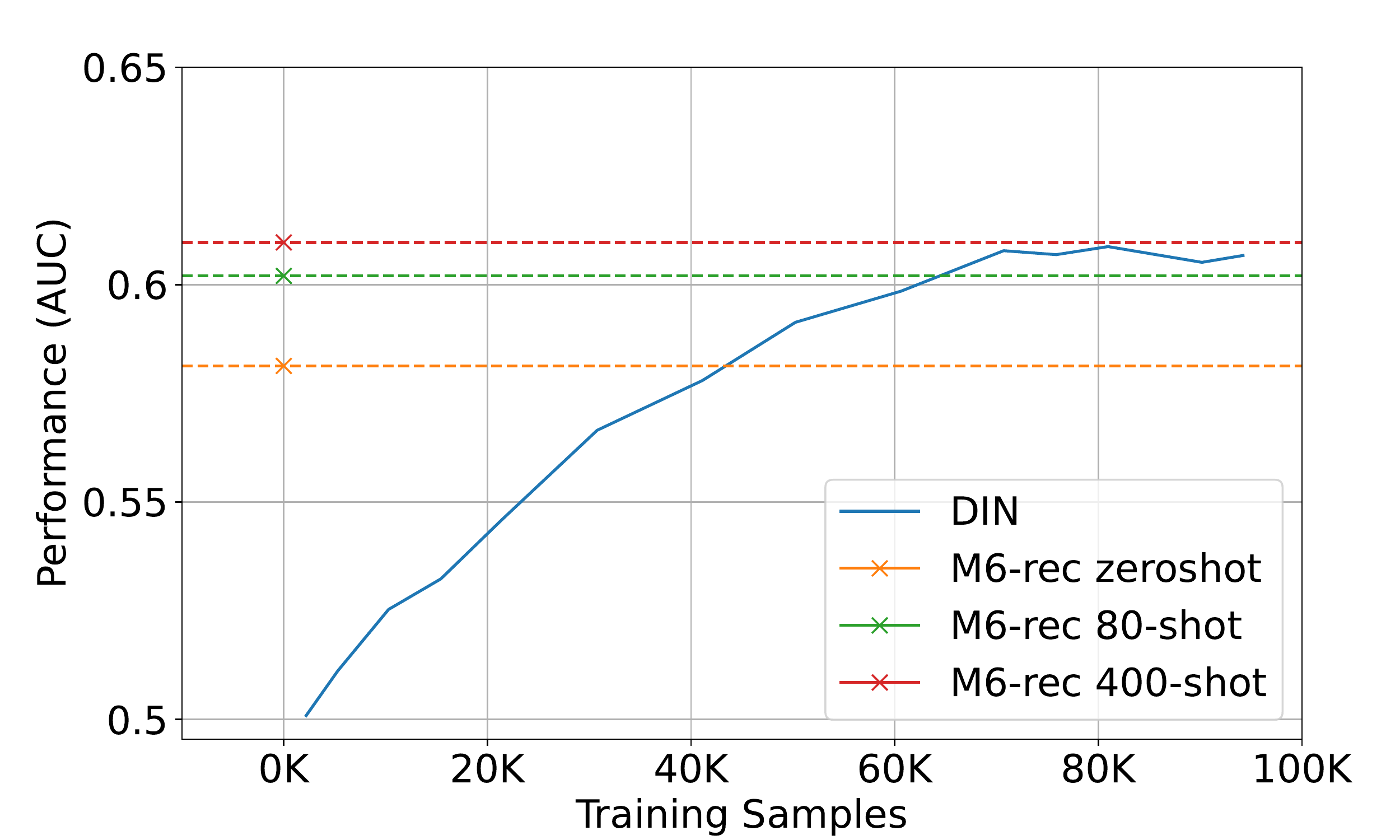}
    }
    \caption{
        The ability to perform zero-shot recommendation. 
    }
    \label{fig:zero-shot-rec}
\end{figure}

\begin{figure}
    \centering
    \includegraphics[width=0.3\textwidth]{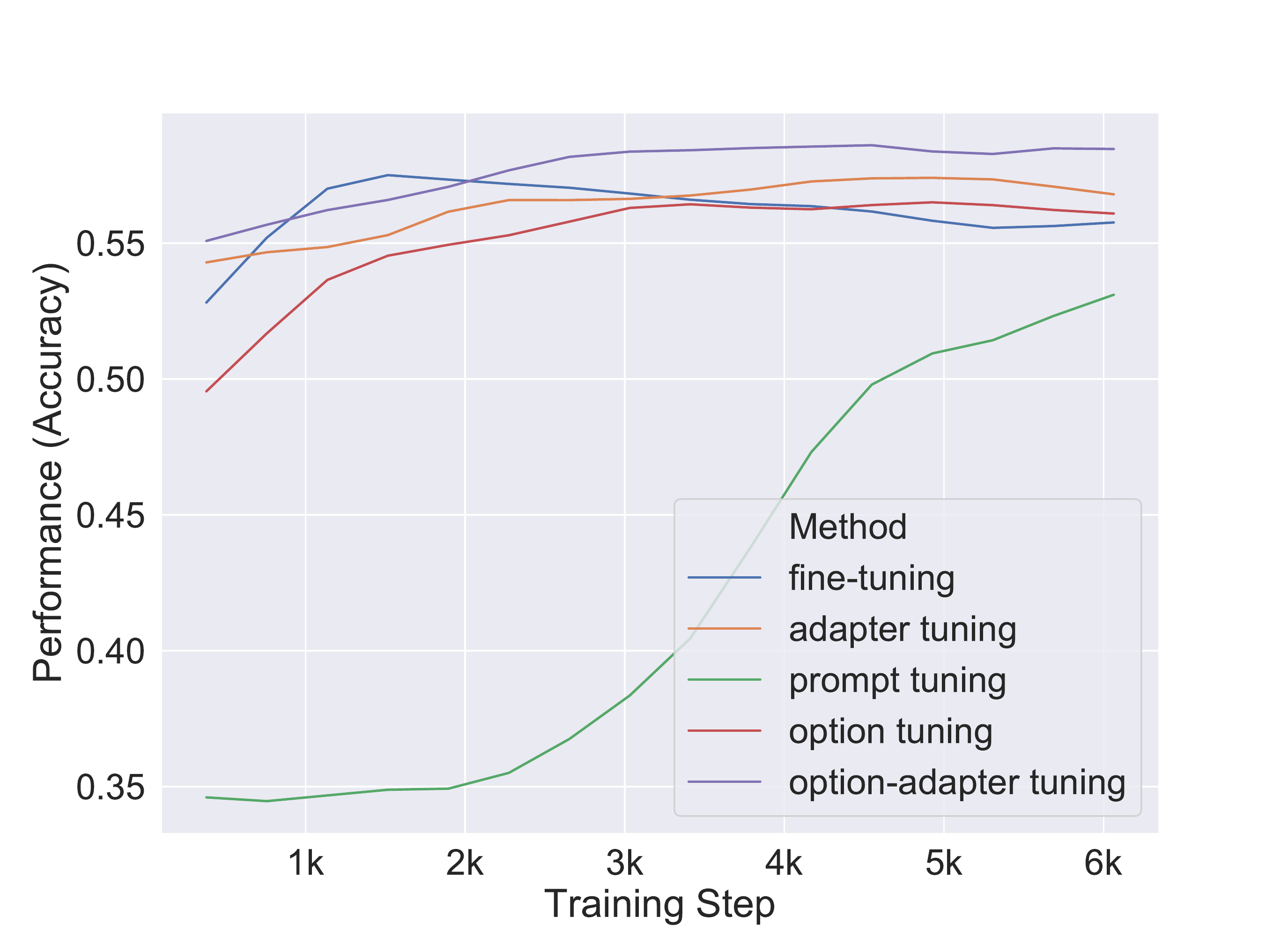}
    \caption{
        Effects of our modifications to prompt tuning, i.e., adding soft options and adapters.
        Option tuning outperforms prompt tuning, and option-adapter tuning outperforms all.
    }
    \label{fig:effect-option-prompt}
\end{figure}

\section{Experiments}\label{sec:exp}

We benchmark M6-Rec on a comprehensive set of tasks, demonstrate its zero-shot learning ability, and conduct ablation studies.

\subsection{Performance on Diverse Tasks}\label{subsec:performance-on-diverse-tasks}

M6-Rec can outperform conventional methods in various stages.

\subsubsection{Ranking}\label{subsubsec:ranking}

Most systems use a deep sequential model such as DIN~\cite{din} for CTR prediction.
DIN stores most features values as integer IDs, learns the embeddings of the IDs, and models the interaction between features via attention modules.
We compare M6-Rec with DIN.
Both methods consider user behavior sequences and use text features such as item titles and category names.
DIN segments each text into phases and stores the phases as integer IDs, while M6-Rec directly operates on the text.
DIN additionally uses item IDs as features, which are widely deemed useful.

Table~\ref{tab:ctr} shows that M6-Rec outperforms DIN on AlipayQuery and TaoProduct, two large-scale datasets collected in Alipay's search query recommender and Taobao's product recommender.
Figure~\ref{fig:ctr-data-scale} further shows that M6-Rec needs less than a million samples to outperform the baseline that requires hundreds of millions of samples.

\subsubsection{Retrieval}\label{subsubsec:retrieval}

We then use M6-Rec for the retrieval (aka.\ candidate generation) of mini-apps in Alipay based on users' locations and past behaviors such as searching and visiting of mini-apps.
We compare M6-Rec with TwinBERT~\cite{twinbert} and YouTubeDNN~\cite{youtubednn}.
TwinBERT fine-tunes BERT for retrieval.
And YouTubeDNN is a traditional baseline that maps feature values to IDs and produces user/item embeddings based on the IDs' embeddings.

Table~\ref{tab:retrieval} shows that M6-Rec outperforms both baselines.
Moreover, when the test cases contain feature values unseen during training, such as new search queries and new mini-apps, the ID-based YouTubeDNN fails, while M6-Rec still performs well.

We also tested M6-Rec in a live experiment, where we use M6-Rec in place of a TwinBERT-like baseline for retrieving mini-apps.
M6-Rec achieved over 1.0\% relative improvement online in terms of CTR and was fully deployed in Alipay's system since July 2021.

\subsubsection{Explanation Generation}\label{subsubsec:explainable-recommendation}

We follow the setting used by PETER~\cite{ACL21-PETER} strictly to measure M6-Rec's performance on generating natural language explanations for justifying the decisions made by the recommender.
Table~\ref{tab:explain-rec} shows that M6-Rec performs far better than the baselines in terms of both explainability and text quality.

\subsubsection{Personalized Content Creation}\label{subsubsec:personalized-content-creation}

We explore two use cases here: (i) generating search queries for recommending to a user, and (ii) generating new product titles based on user behaviors which tell us what types of products may be popular among a user group and help us improve the product supply.

Table~\ref{tab:query-gen} proves M6-Rec's ability to mine useful search queries, following the metrics used by KOBE~\cite{kobe}.
Figure~\ref{fig:tao-product-gen} then showcases some products of the clothing category designed by M6-Rec.
Personalization makes it possible to discover contents catering to users' niche interests and enriches the content supply for recommendation.

\subsubsection{Conversational Recommendation}\label{subsubsec:conversational-recommendation}

We experiment with dialog-based recommendation following the settings of KBRD~\cite{kbrd} on dataset ReDial .
Table~\ref{tab:dialog-rec} shows that M6-Rec makes responses closer to the ground-truth with higher lexical diversity than the baselines.

\subsection{Zero-Shot Recommendation}\label{subsec:zero-shot-recommendation}

The unique advantage of using a pretrained language model as the foundation is that it can judge the likelihood of any event by expressing the event in natural language.
To corroborate this statement, in Figure~\ref{fig:zero-shot-rec} we verify M6-Rec's ability to perform zero-shot ranking on three datasets of different domains, using the zero-shot method described in Subsection~\ref{subsec:beh-as-lang}.
Moreover, after fitting the language loss on a few samples, M6-Rec can match the performance of a traditional ID-based ranker trained on a million samples.

\subsection{Effects of Late Interaction}\label{subsec:effects-of-late-interaction}

Minimizing the latency of real-time inference is critical for some tasks such CTR prediction.
In Table~\ref{tab:late-inter-speed} we report the predictive performance and latency of M6-Rec after applying multi-segment late-interaction.
By caching the immediate results pre-computed by the first $L'=21$ layers, it needs to only compute the last $L-L'=3$ layers when a request arrives.
It thus enjoys a low latency similar to a distilled $3$-layer student model, while incurs much less loss in terms of predictive performance than knowledge distillation.

\subsection{Effects of Option-Adapter Tuning}\label{subsec:effects-of-option-adapter-tuning}

Parameter-efficient tuning is vital to mobile phone deployment, where even one extra MB in the size can harm the user experience and serving many tasks using a shared model is preferred.

In Figure~\ref{fig:effect-option-prompt}, we show that adding soft options to prompt tuning brings significantly better convergence.
And Table~\ref{tab:prompt-tune} shows that the proposed option-adapter tuning is capable of outperforming even full-model fine-tuning, despite tuning only 1\% parameters.

\subsection{Effects of Compression for Mobile Devices}\label{subsec:deployment-on-mobile-devices}

Beside the ``cloud'' rankers on our servers, our systems have additional ``edge'' rankers deployed on the users' mobile phones.
We distill a series of tiny language models named M6-Edge to serve as the foundation for the edge rankers, as described in Subsection~\ref{subsec:model-compress}.

In Table~\ref{tab:edge-model-perf}, we show that M6-Edge outperforms public tiny language models~\footnote{\url{https://github.com/brightmart/albert_zh}} of similar scales on the Chinese benchmark CLUE~\cite{clue-bench}.
An edge ranker built upon M6-Edge has been successfully deployed in Alipay, which brings an around 0.4\% increase in user clicks.

    \section{Conclusion}\label{sec:conclude}

We have proposed M6-Rec, a framework that unifies various tasks in an industrial recommender system, generalizes to open-ended domains, and is capable of performing zero-shot learning.
One future direction is to extend our framework to multimodal settings.

    \bibliographystyle{ACM-Reference-Format}
    \bibliography{main-ref}


\begin{thebibliography}{69}


\ifx \showCODEN    \undefined \def \showCODEN     #1{\unskip}     \fi
\ifx \showDOI      \undefined \def \showDOI       #1{#1}\fi
\ifx \showISBNx    \undefined \def \showISBNx     #1{\unskip}     \fi
\ifx \showISBNxiii \undefined \def \showISBNxiii  #1{\unskip}     \fi
\ifx \showISSN     \undefined \def \showISSN      #1{\unskip}     \fi
\ifx \showLCCN     \undefined \def \showLCCN      #1{\unskip}     \fi
\ifx \shownote     \undefined \def \shownote      #1{#1}          \fi
\ifx \showarticletitle \undefined \def \showarticletitle #1{#1}   \fi
\ifx \showURL      \undefined \def \showURL       {\relax}        \fi
\providecommand\bibfield[2]{#2}
\providecommand\bibinfo[2]{#2}
\providecommand\natexlab[1]{#1}
\providecommand\showeprint[2][]{arXiv:#2}

\bibitem[\protect\citeauthoryear{Bommasani, Hudson, Adeli, Altman, Arora, von
  Arx, Bernstein, Bohg, Bosselut, Brunskill, Brynjolfsson, Buch, Card,
  Castellon, Chatterji, Chen, Creel, Davis, Demszky, Donahue, Doumbouya,
  Durmus, Ermon, Etchemendy, Ethayarajh, Fei{-}Fei, Finn, Gale, Gillespie,
  Goel, Goodman, Grossman, Guha, Hashimoto, Henderson, Hewitt, Ho, Hong, Hsu,
  Huang, Icard, Jain, Jurafsky, Kalluri, Karamcheti, Keeling, Khani, Khattab,
  Koh, Krass, Krishna, Kuditipudi, and et~al.}{Bommasani et~al\mbox{.}}{2021}]%
        {foundation-models}
\bibfield{author}{\bibinfo{person}{Rishi Bommasani}, \bibinfo{person}{Drew~A.
  Hudson}, \bibinfo{person}{Ehsan Adeli}, {et~al\mbox{.}}}
  \bibinfo{year}{2021}\natexlab{}.
\newblock \bibinfo{title}{On the Opportunities and Risks of Foundation Models}.
\newblock
\newblock
\showeprint[arXiv]{2108.07258}


\bibitem[\protect\citeauthoryear{Brown, Mann, Ryder, Subbiah, Kaplan, Dhariwal,
  Neelakantan, Shyam, Sastry, Askell, Agarwal, Herbert{-}Voss, Krueger,
  Henighan, Child, Ramesh, Ziegler, Wu, Winter, Hesse, Chen, Sigler, Litwin,
  Gray, Chess, Clark, Berner, McCandlish, Radford, Sutskever, and Amodei}{Brown
  et~al\mbox{.}}{2020}]%
        {gpt3}
\bibfield{author}{\bibinfo{person}{Tom~B. Brown}, \bibinfo{person}{Benjamin
  Mann}, \bibinfo{person}{Nick Ryder}, {et~al\mbox{.}}}
  \bibinfo{year}{2020}\natexlab{}.
\newblock \showarticletitle{Language Models are Few-Shot Learners}. In
  \bibinfo{booktitle}{\emph{NeurIPS 2020}}.
\newblock


\bibitem[\protect\citeauthoryear{Chen, Lin, Zhang, Ding, Cen, Yang, and
  Tang}{Chen et~al\mbox{.}}{2019a}]%
        {kbrd}
\bibfield{author}{\bibinfo{person}{Qibin Chen}, \bibinfo{person}{Junyang Lin},
  \bibinfo{person}{Yichang Zhang}, {et~al\mbox{.}}}
  \bibinfo{year}{2019}\natexlab{a}.
\newblock \showarticletitle{Towards Knowledge-Based Recommender Dialog System}.
  In \bibinfo{booktitle}{\emph{EMNLP 2019}}.
\newblock


\bibitem[\protect\citeauthoryear{Chen, Lin, Zhang, Yang, Zhou, and Tang}{Chen
  et~al\mbox{.}}{2019b}]%
        {kobe}
\bibfield{author}{\bibinfo{person}{Qibin Chen}, \bibinfo{person}{Junyang Lin},
  \bibinfo{person}{Yichang Zhang}, {et~al\mbox{.}}}
  \bibinfo{year}{2019}\natexlab{b}.
\newblock \showarticletitle{Towards knowledge-based personalized product
  description generation in e-commerce}. In \bibinfo{booktitle}{\emph{KDD
  2019}}.
\newblock


\bibitem[\protect\citeauthoryear{Chen, Frankle, Chang, Liu, Zhang, Wang, and
  Carbin}{Chen et~al\mbox{.}}{2020}]%
        {lotterybert}
\bibfield{author}{\bibinfo{person}{Tianlong Chen}, \bibinfo{person}{Jonathan
  Frankle}, \bibinfo{person}{Shiyu Chang}, {et~al\mbox{.}}}
  \bibinfo{year}{2020}\natexlab{}.
\newblock \showarticletitle{The Lottery Ticket Hypothesis for Pre-trained BERT
  Networks}. In \bibinfo{booktitle}{\emph{NeurIPS 2020}}.
\newblock


\bibitem[\protect\citeauthoryear{Covington, Adams, and Sargin}{Covington
  et~al\mbox{.}}{2016}]%
        {youtubednn}
\bibfield{author}{\bibinfo{person}{Paul Covington}, \bibinfo{person}{Jay
  Adams}, {and} \bibinfo{person}{Emre Sargin}.}
  \bibinfo{year}{2016}\natexlab{}.
\newblock \showarticletitle{Deep Neural Networks for YouTube Recommendations}.
  In \bibinfo{booktitle}{\emph{RecSys 2016}}.
\newblock


\bibitem[\protect\citeauthoryear{Devlin, Chang, Lee, and Toutanova}{Devlin
  et~al\mbox{.}}{2019}]%
        {bert}
\bibfield{author}{\bibinfo{person}{Jacob Devlin}, \bibinfo{person}{Ming-Wei
  Chang}, \bibinfo{person}{Kenton Lee}, {and} \bibinfo{person}{Kristina
  Toutanova}.} \bibinfo{year}{2019}\natexlab{}.
\newblock \showarticletitle{BERT: Pre-training of Deep Bidirectional
  Transformers for Language Understanding}. In \bibinfo{booktitle}{\emph{NAACL
  2019}}.
\newblock


\bibitem[\protect\citeauthoryear{Ding, Yang, Hong, Zheng, Zhou, Yin, Lin, Zou,
  Shao, Yang, and Tang}{Ding et~al\mbox{.}}{2021}]%
        {cogview}
\bibfield{author}{\bibinfo{person}{Ming Ding}, \bibinfo{person}{Zhuoyi Yang},
  \bibinfo{person}{Wenyi Hong}, {et~al\mbox{.}}}
  \bibinfo{year}{2021}\natexlab{}.
\newblock \showarticletitle{CogView: Mastering Text-to-Image Generation via
  Transformers}. In \bibinfo{booktitle}{\emph{NeurIPS 2021}}.
\newblock


\bibitem[\protect\citeauthoryear{Dong, Huang, Wei, Lapata, Zhou, and Xu}{Dong
  et~al\mbox{.}}{2017}]%
        {att2seq}
\bibfield{author}{\bibinfo{person}{Li Dong}, \bibinfo{person}{Shaohan Huang},
  \bibinfo{person}{Furu Wei}, {et~al\mbox{.}}} \bibinfo{year}{2017}\natexlab{}.
\newblock \showarticletitle{Learning to Generate Product Reviews from
  Attributes}. In \bibinfo{booktitle}{\emph{EACL 2017}}.
\newblock


\bibitem[\protect\citeauthoryear{Dong, Yang, Wang, Wei, Liu, Wang, Gao, Zhou,
  and Hon}{Dong et~al\mbox{.}}{2019}]%
        {unilm}
\bibfield{author}{\bibinfo{person}{Li Dong}, \bibinfo{person}{Nan Yang},
  \bibinfo{person}{Wenhui Wang}, {et~al\mbox{.}}}
  \bibinfo{year}{2019}\natexlab{}.
\newblock \showarticletitle{Unified Language Model Pre-training for Natural
  Language Understanding and Generation}. In \bibinfo{booktitle}{\emph{NeurIPS
  2019}}.
\newblock


\bibitem[\protect\citeauthoryear{Fedus, Zoph, and Shazeer}{Fedus
  et~al\mbox{.}}{2021}]%
        {switchtrm}
\bibfield{author}{\bibinfo{person}{William Fedus}, \bibinfo{person}{Barret
  Zoph}, {and} \bibinfo{person}{Noam Shazeer}.}
  \bibinfo{year}{2021}\natexlab{}.
\newblock \bibinfo{title}{Switch Transformers: Scaling to Trillion Parameter
  Models with Simple and Efficient Sparsity}.
\newblock
\newblock


\bibitem[\protect\citeauthoryear{Gordon, Duh, and Andrews}{Gordon
  et~al\mbox{.}}{2020}]%
        {compress_bert}
\bibfield{author}{\bibinfo{person}{Mitchell~A Gordon}, \bibinfo{person}{Kevin
  Duh}, {and} \bibinfo{person}{Nicholas Andrews}.}
  \bibinfo{year}{2020}\natexlab{}.
\newblock \showarticletitle{Compressing BERT: Studying the Effects of Weight
  Pruning on Transfer Learning}. In \bibinfo{booktitle}{\emph{ICLR 2020}}.
\newblock


\bibitem[\protect\citeauthoryear{He, Zhou, Ma, Berg{-}Kirkpatrick, and
  Neubig}{He et~al\mbox{.}}{2022}]%
        {unify-eff-tune}
\bibfield{author}{\bibinfo{person}{Junxian He}, \bibinfo{person}{Chunting
  Zhou}, \bibinfo{person}{Xuezhe Ma}, \bibinfo{person}{Taylor
  Berg{-}Kirkpatrick}, {and} \bibinfo{person}{Graham Neubig}.}
  \bibinfo{year}{2022}\natexlab{}.
\newblock \showarticletitle{Towards a Unified View of Parameter-Efficient
  Transfer Learning}. In \bibinfo{booktitle}{\emph{ICLR 2022}}.
\newblock


\bibitem[\protect\citeauthoryear{Hou, Huang, Shang, Jiang, Chen, and Liu}{Hou
  et~al\mbox{.}}{2020}]%
        {dynabert}
\bibfield{author}{\bibinfo{person}{Lu Hou}, \bibinfo{person}{Zhiqi Huang},
  \bibinfo{person}{Lifeng Shang}, {et~al\mbox{.}}}
  \bibinfo{year}{2020}\natexlab{}.
\newblock \showarticletitle{DynaBERT: Dynamic BERT with Adaptive Width and
  Depth}. In \bibinfo{booktitle}{\emph{NeurIPS 2020}}.
\newblock


\bibitem[\protect\citeauthoryear{Hu, yelong shen, Wallis, Allen-Zhu, Li, Wang,
  Wang, and Chen}{Hu et~al\mbox{.}}{2022}]%
        {lora_adapter}
\bibfield{author}{\bibinfo{person}{Edward~J Hu}, \bibinfo{person}{yelong shen},
  \bibinfo{person}{Phillip Wallis}, {et~al\mbox{.}}}
  \bibinfo{year}{2022}\natexlab{}.
\newblock \showarticletitle{LoRA: Low-Rank Adaptation of Large Language
  Models}. In \bibinfo{booktitle}{\emph{ICLR 2022}}.
\newblock


\bibitem[\protect\citeauthoryear{Huang, Chen, Li, Wu, van~der Maaten, and
  Weinberger}{Huang et~al\mbox{.}}{2018}]%
        {msdnet}
\bibfield{author}{\bibinfo{person}{Gao Huang}, \bibinfo{person}{Danlu Chen},
  \bibinfo{person}{Tianhong Li}, {et~al\mbox{.}}}
  \bibinfo{year}{2018}\natexlab{}.
\newblock \showarticletitle{Multi-Scale Dense Networks for Resource Efficient
  Image Classification}. In \bibinfo{booktitle}{\emph{ICLR 2018}}.
\newblock


\bibitem[\protect\citeauthoryear{Jiao, Yin, Shang, Jiang, Chen, Li, Wang, and
  Liu}{Jiao et~al\mbox{.}}{2020}]%
        {tinybert}
\bibfield{author}{\bibinfo{person}{Xiaoqi Jiao}, \bibinfo{person}{Yichun Yin},
  \bibinfo{person}{Lifeng Shang}, {et~al\mbox{.}}}
  \bibinfo{year}{2020}\natexlab{}.
\newblock \showarticletitle{TinyBERT: Distilling BERT for Natural Language
  Understanding}. In \bibinfo{booktitle}{\emph{Findings of EMNLP 2020}}.
\newblock


\bibitem[\protect\citeauthoryear{Kang, Fang, Wang, and McAuley}{Kang
  et~al\mbox{.}}{2017}]%
        {kang17visually}
\bibfield{author}{\bibinfo{person}{Wang-Cheng Kang}, \bibinfo{person}{Chen
  Fang}, \bibinfo{person}{Zhaowen Wang}, {and} \bibinfo{person}{Julian
  McAuley}.} \bibinfo{year}{2017}\natexlab{}.
\newblock \showarticletitle{Visually-aware fashion recommendation and design
  with generative image models}. In \bibinfo{booktitle}{\emph{ICDM 2017}}.
\newblock


\bibitem[\protect\citeauthoryear{Khattab and Zaharia}{Khattab and
  Zaharia}{2020}]%
        {colbert}
\bibfield{author}{\bibinfo{person}{Omar Khattab} {and} \bibinfo{person}{Matei
  Zaharia}.} \bibinfo{year}{2020}\natexlab{}.
\newblock \showarticletitle{ColBERT: Efficient and Effective Passage Search via
  Contextualized Late Interaction over BERT}. In
  \bibinfo{booktitle}{\emph{SIGIR 2020}}.
\newblock


\bibitem[\protect\citeauthoryear{Lan, Chen, Goodman, Gimpel, Sharma, and
  Soricut}{Lan et~al\mbox{.}}{2020}]%
        {albert}
\bibfield{author}{\bibinfo{person}{Zhenzhong Lan}, \bibinfo{person}{Mingda
  Chen}, \bibinfo{person}{Sebastian Goodman}, {et~al\mbox{.}}}
  \bibinfo{year}{2020}\natexlab{}.
\newblock \showarticletitle{ALBERT: A Lite BERT for Self-supervised Learning of
  Language Representations}. In \bibinfo{booktitle}{\emph{ICLR 2020}}.
\newblock


\bibitem[\protect\citeauthoryear{Lepikhin, Lee, Xu, Chen, Firat, Huang, Krikun,
  Shazeer, and Chen}{Lepikhin et~al\mbox{.}}{2021}]%
        {gshard}
\bibfield{author}{\bibinfo{person}{Dmitry Lepikhin},
  \bibinfo{person}{HyoukJoong Lee}, \bibinfo{person}{Yuanzhong Xu},
  {et~al\mbox{.}}} \bibinfo{year}{2021}\natexlab{}.
\newblock \showarticletitle{GShard: Scaling Giant Models with Conditional
  Computation and Automatic Sharding}. In \bibinfo{booktitle}{\emph{ICLR}}.
\newblock


\bibitem[\protect\citeauthoryear{Lester, Al-Rfou, and Constant}{Lester
  et~al\mbox{.}}{2021}]%
        {prompt_tune}
\bibfield{author}{\bibinfo{person}{Brian Lester}, \bibinfo{person}{Rami
  Al-Rfou}, {and} \bibinfo{person}{Noah Constant}.}
  \bibinfo{year}{2021}\natexlab{}.
\newblock \showarticletitle{The Power of Scale for Parameter-Efficient Prompt
  Tuning}. In \bibinfo{booktitle}{\emph{EMNLP 2021}}.
\newblock


\bibitem[\protect\citeauthoryear{Lewis, Liu, Goyal, Ghazvininejad, Mohamed,
  Levy, Stoyanov, and Zettlemoyer}{Lewis et~al\mbox{.}}{2020}]%
        {bart}
\bibfield{author}{\bibinfo{person}{Mike Lewis}, \bibinfo{person}{Yinhan Liu},
  \bibinfo{person}{Naman Goyal}, {et~al\mbox{.}}}
  \bibinfo{year}{2020}\natexlab{}.
\newblock \showarticletitle{BART: Denoising Sequence-to-Sequence Pre-training
  for Natural Language Generation, Translation, and Comprehension}. In
  \bibinfo{booktitle}{\emph{ACL 2020}}.
\newblock


\bibitem[\protect\citeauthoryear{Li, Zhang, and Chen}{Li et~al\mbox{.}}{2020}]%
        {CIKM20-NETE}
\bibfield{author}{\bibinfo{person}{Lei Li}, \bibinfo{person}{Yongfeng Zhang},
  {and} \bibinfo{person}{Li Chen}.} \bibinfo{year}{2020}\natexlab{}.
\newblock \showarticletitle{Generate Neural Template Explanations for
  Recommendation}. In \bibinfo{booktitle}{\emph{CIKM 2020}}.
\newblock


\bibitem[\protect\citeauthoryear{Li, Zhang, and Chen}{Li et~al\mbox{.}}{2021}]%
        {ACL21-PETER}
\bibfield{author}{\bibinfo{person}{Lei Li}, \bibinfo{person}{Yongfeng Zhang},
  {and} \bibinfo{person}{Li Chen}.} \bibinfo{year}{2021}\natexlab{}.
\newblock \showarticletitle{Personalized Transformer for Explainable
  Recommendation}. In \bibinfo{booktitle}{\emph{ACL 2021}}.
\newblock


\bibitem[\protect\citeauthoryear{Li, Wang, Ren, Bing, and Lam}{Li
  et~al\mbox{.}}{2017}]%
        {nrt}
\bibfield{author}{\bibinfo{person}{Piji Li}, \bibinfo{person}{Zihao Wang},
  \bibinfo{person}{Zhaochun Ren}, \bibinfo{person}{Lidong Bing}, {and}
  \bibinfo{person}{Wai Lam}.} \bibinfo{year}{2017}\natexlab{}.
\newblock \showarticletitle{Neural Rating Regression with Abstractive Tips
  Generation for Recommendation}. In \bibinfo{booktitle}{\emph{SIGIR 2017}}.
\newblock


\bibitem[\protect\citeauthoryear{Li and Liang}{Li and Liang}{2021}]%
        {prefix_tune}
\bibfield{author}{\bibinfo{person}{Xiang~Lisa Li} {and} \bibinfo{person}{Percy
  Liang}.} \bibinfo{year}{2021}\natexlab{}.
\newblock \showarticletitle{Prefix-Tuning: Optimizing Continuous Prompts for
  Generation}. In \bibinfo{booktitle}{\emph{ACL 2021}}.
\newblock


\bibitem[\protect\citeauthoryear{Lin, Men, Yang, Zhou, Ding, Zhang, Wang, Wang,
  Jiang, Jia, Zhang, Zhang, Zou, Li, Deng, Liu, Xue, Zhou, Ma, Yu, Li, Lin,
  Zhou, Tang, and Yang}{Lin et~al\mbox{.}}{2021a}]%
        {m6-arxiv}
\bibfield{author}{\bibinfo{person}{Junyang Lin}, \bibinfo{person}{Rui Men},
  \bibinfo{person}{An Yang}, {et~al\mbox{.}}} \bibinfo{year}{2021}\natexlab{a}.
\newblock \bibinfo{title}{M6: A Chinese Multimodal Pretrainer}.
\newblock
\newblock
\showeprint[arXiv]{2103.00823}


\bibitem[\protect\citeauthoryear{Lin, Men, Yang, Zhou, Zhang, Wang, Zhou, Tang,
  and Yang}{Lin et~al\mbox{.}}{2021b}]%
        {m6-kdd}
\bibfield{author}{\bibinfo{person}{Junyang Lin}, \bibinfo{person}{Rui Men},
  \bibinfo{person}{An Yang}, {et~al\mbox{.}}} \bibinfo{year}{2021}\natexlab{b}.
\newblock \showarticletitle{M6: Multi-Modality-to-Multi-Modality Multitask
  Mega-transformer for Unified Pretraining}. In \bibinfo{booktitle}{\emph{KDD
  2021}}.
\newblock


\bibitem[\protect\citeauthoryear{Lin, Yang, Bai, Zhou, Jiang, Jia, Wang, Zhang,
  Li, Lin, Zhou, and Yang}{Lin et~al\mbox{.}}{2021c}]%
        {m6-10t}
\bibfield{author}{\bibinfo{person}{Junyang Lin}, \bibinfo{person}{An Yang},
  \bibinfo{person}{Jinze Bai}, {et~al\mbox{.}}}
  \bibinfo{year}{2021}\natexlab{c}.
\newblock \bibinfo{title}{M6-10T: A Sharing-Delinking Paradigm for Efficient
  Multi-Trillion Parameter Pretraining}.
\newblock
\newblock
\showeprint[arXiv]{2110.03888}


\bibitem[\protect\citeauthoryear{Liu, Huang, Liu, Lu, Cheng, Li, Shi, Wang,
  Cheng, and Yin}{Liu et~al\mbox{.}}{2021}]%
        {ernie-retrieval}
\bibfield{author}{\bibinfo{person}{Yiding Liu}, \bibinfo{person}{Guan Huang},
  \bibinfo{person}{Jiaxiang Liu}, {et~al\mbox{.}}}
  \bibinfo{year}{2021}\natexlab{}.
\newblock \showarticletitle{Pre-trained Language Model for Web-scale Retrieval
  in Baidu Search}. In \bibinfo{booktitle}{\emph{KDD 2021}}.
\newblock


\bibitem[\protect\citeauthoryear{Lu, Batra, Parikh, and Lee}{Lu
  et~al\mbox{.}}{2019}]%
        {vilbert}
\bibfield{author}{\bibinfo{person}{Jiasen Lu}, \bibinfo{person}{Dhruv Batra},
  \bibinfo{person}{Devi Parikh}, {and} \bibinfo{person}{Stefan Lee}.}
  \bibinfo{year}{2019}\natexlab{}.
\newblock \showarticletitle{ViLBERT: Pretraining Task-Agnostic Visiolinguistic
  Representations for Vision-and-Language Tasks}. In
  \bibinfo{booktitle}{\emph{NeurIPS 2019}}.
\newblock


\bibitem[\protect\citeauthoryear{Lu, Jiao, and Zhang}{Lu et~al\mbox{.}}{2020}]%
        {twinbert}
\bibfield{author}{\bibinfo{person}{Wenhao Lu}, \bibinfo{person}{Jian Jiao},
  {and} \bibinfo{person}{Ruofei Zhang}.} \bibinfo{year}{2020}\natexlab{}.
\newblock \bibinfo{title}{TwinBERT: Distilling Knowledge to Twin-Structured
  BERT Models for Efficient Retrieval}.
\newblock
\newblock
\showeprint[arXiv]{2002.06275}


\bibitem[\protect\citeauthoryear{Nakano, Hilton, Balaji, Wu, Ouyang, Kim,
  Hesse, Jain, Kosaraju, Saunders, Jiang, Cobbe, Eloundou, Krueger, Button,
  Knight, Chess, and Schulman}{Nakano et~al\mbox{.}}{2021}]%
        {webgpt}
\bibfield{author}{\bibinfo{person}{Reiichiro Nakano}, \bibinfo{person}{Jacob
  Hilton}, \bibinfo{person}{Suchir Balaji}, {et~al\mbox{.}}}
  \bibinfo{year}{2021}\natexlab{}.
\newblock \bibinfo{title}{WebGPT: Browser-assisted question-answering with
  human feedback}.
\newblock
\newblock
\showeprint[arXiv]{2112.09332}


\bibitem[\protect\citeauthoryear{Ni, Li, and McAuley}{Ni
  et~al\mbox{.}}{2019a}]%
        {ACMLM}
\bibfield{author}{\bibinfo{person}{Jianmo Ni}, \bibinfo{person}{Jiacheng Li},
  {and} \bibinfo{person}{Julian McAuley}.} \bibinfo{year}{2019}\natexlab{a}.
\newblock \showarticletitle{Justifying Recommendations using Distantly-Labeled
  Reviews and Fine-Grained Aspects}. In \bibinfo{booktitle}{\emph{EMNLP 2019}}.
\newblock


\bibitem[\protect\citeauthoryear{Ni, Li, and McAuley}{Ni
  et~al\mbox{.}}{2019b}]%
        {amazon-data-2019}
\bibfield{author}{\bibinfo{person}{Jianmo Ni}, \bibinfo{person}{Jiacheng Li},
  {and} \bibinfo{person}{Julian McAuley}.} \bibinfo{year}{2019}\natexlab{b}.
\newblock \showarticletitle{Justifying Recommendations using Distantly-Labeled
  Reviews and Fine-Grained Aspects}. In \bibinfo{booktitle}{\emph{EMNLP 2019}}.
\newblock


\bibitem[\protect\citeauthoryear{Radford, Kim, Hallacy, Ramesh, Goh, Agarwal,
  Sastry, Askell, Mishkin, Clark, Krueger, and Sutskever}{Radford
  et~al\mbox{.}}{2021}]%
        {clip}
\bibfield{author}{\bibinfo{person}{Alec Radford}, \bibinfo{person}{Jong~Wook
  Kim}, \bibinfo{person}{Chris Hallacy}, {et~al\mbox{.}}}
  \bibinfo{year}{2021}\natexlab{}.
\newblock \showarticletitle{Learning Transferable Visual Models From Natural
  Language Supervision}. In \bibinfo{booktitle}{\emph{ICML 2021}}.
\newblock


\bibitem[\protect\citeauthoryear{Radford, Wu, Child, Luan, Amodei, and
  Sutskever}{Radford et~al\mbox{.}}{2019}]%
        {gpt2}
\bibfield{author}{\bibinfo{person}{Alec Radford}, \bibinfo{person}{Jeff Wu},
  \bibinfo{person}{Rewon Child}, {et~al\mbox{.}}}
  \bibinfo{year}{2019}\natexlab{}.
\newblock \bibinfo{title}{Language Models are Unsupervised Multitask Learners}.
\newblock
\newblock


\bibitem[\protect\citeauthoryear{Raffel, Shazeer, Roberts, Lee, Narang, Matena,
  Zhou, Li, and Liu}{Raffel et~al\mbox{.}}{2020}]%
        {2020t5}
\bibfield{author}{\bibinfo{person}{Colin Raffel}, \bibinfo{person}{Noam
  Shazeer}, \bibinfo{person}{Adam Roberts}, {et~al\mbox{.}}}
  \bibinfo{year}{2020}\natexlab{}.
\newblock \showarticletitle{Exploring the Limits of Transfer Learning with a
  Unified Text-to-Text Transformer}.
\newblock \bibinfo{journal}{\emph{JMLR}} (\bibinfo{year}{2020}).
\newblock


\bibitem[\protect\citeauthoryear{Ramesh, Pavlov, Goh, Gray, Voss, Radford,
  Chen, and Sutskever}{Ramesh et~al\mbox{.}}{2021}]%
        {dall-e}
\bibfield{author}{\bibinfo{person}{Aditya Ramesh}, \bibinfo{person}{Mikhail
  Pavlov}, \bibinfo{person}{Gabriel Goh}, {et~al\mbox{.}}}
  \bibinfo{year}{2021}\natexlab{}.
\newblock \showarticletitle{Zero-Shot Text-to-Image Generation}. In
  \bibinfo{booktitle}{\emph{ICML 2021}}.
\newblock


\bibitem[\protect\citeauthoryear{Sanh, Debut, Chaumond, and Wolf}{Sanh
  et~al\mbox{.}}{2019}]%
        {distillbert}
\bibfield{author}{\bibinfo{person}{Victor Sanh}, \bibinfo{person}{Lysandre
  Debut}, \bibinfo{person}{Julien Chaumond}, {and} \bibinfo{person}{Thomas
  Wolf}.} \bibinfo{year}{2019}\natexlab{}.
\newblock \bibinfo{title}{DistilBERT, a distilled version of BERT: smaller,
  faster, cheaper and lighter}.
\newblock
\newblock
\showeprint[arXiv]{1910.01108}


\bibitem[\protect\citeauthoryear{Shoeybi, Patwary, Puri, LeGresley, Casper, and
  Catanzaro}{Shoeybi et~al\mbox{.}}{2019}]%
        {megatron-lm}
\bibfield{author}{\bibinfo{person}{Mohammad Shoeybi}, \bibinfo{person}{Mostofa
  Patwary}, \bibinfo{person}{Raul Puri}, {et~al\mbox{.}}}
  \bibinfo{year}{2019}\natexlab{}.
\newblock \bibinfo{title}{Megatron-LM: Training Multi-Billion Parameter
  Language Models Using Model Parallelism}.
\newblock
\newblock
\showeprint[arXiv]{1909.08053}


\bibitem[\protect\citeauthoryear{Sileo, Vossen, and Raymaekers}{Sileo
  et~al\mbox{.}}{2022}]%
        {zero-rec-lang}
\bibfield{author}{\bibinfo{person}{Damien Sileo}, \bibinfo{person}{Wout
  Vossen}, {and} \bibinfo{person}{Robbe Raymaekers}.}
  \bibinfo{year}{2022}\natexlab{}.
\newblock \showarticletitle{Zero-Shot Recommendation as Language Modeling}. In
  \bibinfo{booktitle}{\emph{ECIR 2022}}.
\newblock


\bibitem[\protect\citeauthoryear{Su, Zhu, Cao, Li, Lu, Wei, and Dai}{Su
  et~al\mbox{.}}{2019}]%
        {vl-bert}
\bibfield{author}{\bibinfo{person}{Weijie Su}, \bibinfo{person}{Xizhou Zhu},
  \bibinfo{person}{Yue Cao}, {et~al\mbox{.}}} \bibinfo{year}{2019}\natexlab{}.
\newblock \showarticletitle{VL-BERT: Pre-training of Generic Visual-Linguistic
  Representations}. In \bibinfo{booktitle}{\emph{ICLR 2020}}.
\newblock


\bibitem[\protect\citeauthoryear{Su, Wang, Qin, Chan, Lin, Liu, Li, Li, Hou,
  Sun, and Zhou}{Su et~al\mbox{.}}{2021}]%
        {transfer_prompt}
\bibfield{author}{\bibinfo{person}{YuSheng Su}, \bibinfo{person}{Xiaozhi Wang},
  \bibinfo{person}{Yujia Qin}, {et~al\mbox{.}}}
  \bibinfo{year}{2021}\natexlab{}.
\newblock \bibinfo{title}{On Transferability of Prompt Tuning for Natural
  Language Understanding}.
\newblock
\newblock
\showeprint[arXiv]{2111.06719}


\bibitem[\protect\citeauthoryear{Sun, Wang, Feng, Ding, Pang, Shang, Liu, Chen,
  Zhao, Lu, Liu, Wu, Gong, Liang, Shang, Sun, Liu, Ouyang, Yu, Tian, Wu, and
  Wang}{Sun et~al\mbox{.}}{2021}]%
        {ernie3}
\bibfield{author}{\bibinfo{person}{Yu Sun}, \bibinfo{person}{Shuohuan Wang},
  \bibinfo{person}{Shikun Feng}, {et~al\mbox{.}}}
  \bibinfo{year}{2021}\natexlab{}.
\newblock \bibinfo{title}{ERNIE 3.0: Large-scale Knowledge Enhanced
  Pre-training for Language Understanding and Generation}.
\newblock
\newblock
\showeprint[arXiv]{2107.02137}


\bibitem[\protect\citeauthoryear{Sun, Yu, Song, Liu, Yang, and Zhou}{Sun
  et~al\mbox{.}}{2020}]%
        {mobilebert}
\bibfield{author}{\bibinfo{person}{Zhiqing Sun}, \bibinfo{person}{Hongkun Yu},
  \bibinfo{person}{Xiaodan Song}, {et~al\mbox{.}}}
  \bibinfo{year}{2020}\natexlab{}.
\newblock \showarticletitle{MobileBERT: a Compact Task-Agnostic BERT for
  Resource-Limited Devices}. In \bibinfo{booktitle}{\emph{ACL 2020}}.
\newblock


\bibitem[\protect\citeauthoryear{Teerapittayanon, McDanel, and
  Kung}{Teerapittayanon et~al\mbox{.}}{2016}]%
        {branchynet}
\bibfield{author}{\bibinfo{person}{Surat Teerapittayanon},
  \bibinfo{person}{Bradley McDanel}, {and} \bibinfo{person}{H.~T. Kung}.}
  \bibinfo{year}{2016}\natexlab{}.
\newblock \showarticletitle{BranchyNet: Fast Inference via Early Exiting from
  Deep Neural Networks}. In \bibinfo{booktitle}{\emph{ICPR 2016}}.
\newblock


\bibitem[\protect\citeauthoryear{Vaswani, Shazeer, Parmar, Uszkoreit, Jones,
  Gomez, Kaiser, and Polosukhin}{Vaswani et~al\mbox{.}}{2017}]%
        {transformer}
\bibfield{author}{\bibinfo{person}{Ashish Vaswani}, \bibinfo{person}{Noam
  Shazeer}, \bibinfo{person}{Niki Parmar}, {et~al\mbox{.}}}
  \bibinfo{year}{2017}\natexlab{}.
\newblock \showarticletitle{Attention is All you Need}. In
  \bibinfo{booktitle}{\emph{NeurIPS 2017}}.
\newblock


\bibitem[\protect\citeauthoryear{Wang, Wei, Dong, Bao, Yang, and Zhou}{Wang
  et~al\mbox{.}}{2020}]%
        {minilm}
\bibfield{author}{\bibinfo{person}{Wenhui Wang}, \bibinfo{person}{Furu Wei},
  \bibinfo{person}{Li Dong}, {et~al\mbox{.}}} \bibinfo{year}{2020}\natexlab{}.
\newblock \showarticletitle{MiniLM: Deep Self-Attention Distillation for
  Task-Agnostic Compression of Pre-Trained Transformers}. In
  \bibinfo{booktitle}{\emph{NeurIPS 2020}}.
\newblock


\bibitem[\protect\citeauthoryear{Wang, Thain, Sinha, Prost, Chi, Chen, and
  Beutel}{Wang et~al\mbox{.}}{2021}]%
        {fair-multi-recsys}
\bibfield{author}{\bibinfo{person}{Xuezhi Wang}, \bibinfo{person}{Nithum
  Thain}, \bibinfo{person}{Anu~Aradhana Sinha}, {et~al\mbox{.}}}
  \bibinfo{year}{2021}\natexlab{}.
\newblock \showarticletitle{Practical Compositional Fairness: Understanding
  Fairness in Multi-Component Recommender Systems}. In
  \bibinfo{booktitle}{\emph{WSDM 2021}}.
\newblock


\bibitem[\protect\citeauthoryear{Wu, Wu, Qi, Lian, Huang, and Xie}{Wu
  et~al\mbox{.}}{2020}]%
        {ptum}
\bibfield{author}{\bibinfo{person}{Chuhan Wu}, \bibinfo{person}{Fangzhao Wu},
  \bibinfo{person}{Tao Qi}, {et~al\mbox{.}}} \bibinfo{year}{2020}\natexlab{}.
\newblock \showarticletitle{PTUM: Pre-training User Model from Unlabeled User
  Behaviors via Self-supervision}. In \bibinfo{booktitle}{\emph{Findings of
  EMNLP 2020}}.
\newblock


\bibitem[\protect\citeauthoryear{Xin, Tang, Lee, Yu, and Lin}{Xin
  et~al\mbox{.}}{2020}]%
        {deebert}
\bibfield{author}{\bibinfo{person}{Ji Xin}, \bibinfo{person}{Raphael Tang},
  \bibinfo{person}{Jaejun Lee}, \bibinfo{person}{Yaoliang Yu}, {and}
  \bibinfo{person}{Jimmy Lin}.} \bibinfo{year}{2020}\natexlab{}.
\newblock \showarticletitle{DeeBERT: Dynamic Early Exiting for Accelerating
  BERT Inference}. In \bibinfo{booktitle}{\emph{ACL 2020}}.
\newblock


\bibitem[\protect\citeauthoryear{Xu, Hu, Zhang, Li, Cao, Li, Xu, Sun, Yu, Yu,
  Tian, Dong, Liu, Shi, Cui, Li, Zeng, Wang, Xie, Li, Patterson, Tian, Zhang,
  Zhou, Liu, Zhao, Zhao, Yue, Zhang, Yang, Richardson, and Lan}{Xu
  et~al\mbox{.}}{2020}]%
        {clue-bench}
\bibfield{author}{\bibinfo{person}{Liang Xu}, \bibinfo{person}{Hai Hu},
  \bibinfo{person}{Xuanwei Zhang}, {et~al\mbox{.}}}
  \bibinfo{year}{2020}\natexlab{}.
\newblock \showarticletitle{CLUE: A Chinese Language Understanding Evaluation
  Benchmark}. In \bibinfo{booktitle}{\emph{COLING 2020}}.
\newblock


\bibitem[\protect\citeauthoryear{Yuan, He, Karatzoglou, and Zhang}{Yuan
  et~al\mbox{.}}{2020}]%
        {yuan2020parameter}
\bibfield{author}{\bibinfo{person}{Fajie Yuan}, \bibinfo{person}{Xiangnan He},
  \bibinfo{person}{Alexandros Karatzoglou}, {and} \bibinfo{person}{Liguang
  Zhang}.} \bibinfo{year}{2020}\natexlab{}.
\newblock \showarticletitle{Parameter-Efficient Transfer from Sequential
  Behaviors for User Modeling and Recommendation}.
\newblock \bibinfo{journal}{\emph{SIGIR 2020}}.
\newblock


\bibitem[\protect\citeauthoryear{Yuan, Zhang, Karatzoglou, Jose, Kong, and
  Li}{Yuan et~al\mbox{.}}{2021}]%
        {yuan2021one}
\bibfield{author}{\bibinfo{person}{Fajie Yuan}, \bibinfo{person}{Guoxiao
  Zhang}, \bibinfo{person}{Alexandros Karatzoglou}, {et~al\mbox{.}}}
  \bibinfo{year}{2021}\natexlab{}.
\newblock \showarticletitle{One person, one model, one world: Learning
  continual user representation without forgetting}. In
  \bibinfo{booktitle}{\emph{SIGIR 2021}}.
\newblock


\bibitem[\protect\citeauthoryear{Zafrir, Boudoukh, Izsak, and
  Wasserblat}{Zafrir et~al\mbox{.}}{2019}]%
        {q8bert}
\bibfield{author}{\bibinfo{person}{Ofir Zafrir}, \bibinfo{person}{Guy
  Boudoukh}, \bibinfo{person}{Peter Izsak}, {and} \bibinfo{person}{Moshe
  Wasserblat}.} \bibinfo{year}{2019}\natexlab{}.
\newblock \bibinfo{title}{Q8BERT: Quantized 8Bit BERT}.
\newblock
\newblock
\showeprint[arXiv]{1910.06188}


\bibitem[\protect\citeauthoryear{Zafrir, Larey, Boudoukh, Shen, and
  Wasserblat}{Zafrir et~al\mbox{.}}{2021}]%
        {pofa_bert}
\bibfield{author}{\bibinfo{person}{Ofir Zafrir}, \bibinfo{person}{Ariel Larey},
  \bibinfo{person}{Guy Boudoukh}, \bibinfo{person}{Haihao Shen}, {and}
  \bibinfo{person}{Moshe Wasserblat}.} \bibinfo{year}{2021}\natexlab{}.
\newblock \bibinfo{title}{Prune Once for All: Sparse Pre-Trained Language
  Models}.
\newblock
\newblock
\showeprint[arXiv]{2111.05754}


\bibitem[\protect\citeauthoryear{Zeng, Ren, Su, Wang, Liao, Wang, Jiang, Yang,
  Wang, Zhang, Li, Gong, Yao, Huang, Wang, Yu, Guo, Yu, Zhang, Wang, Tao, Yan,
  Yi, Peng, Jiang, Zhang, Deng, Zhang, Lin, Zhang, Zhang, Guo, Gu, Fan, Wang,
  Jin, Liu, and Tian}{Zeng et~al\mbox{.}}{2021a}]%
        {pangu-alpha}
\bibfield{author}{\bibinfo{person}{Wei Zeng}, \bibinfo{person}{Xiaozhe Ren},
  \bibinfo{person}{Teng Su}, {et~al\mbox{.}}} \bibinfo{year}{2021}\natexlab{a}.
\newblock \bibinfo{title}{PanGu-$\alpha$: Large-scale Autoregressive Pretrained
  Chinese Language Models with Auto-parallel Computation}.
\newblock
\newblock
\showeprint[arXiv]{2104.12369}


\bibitem[\protect\citeauthoryear{Zeng, Xiao, Yao, Xie, Liu, Lin, Lin, and
  Sun}{Zeng et~al\mbox{.}}{2021b}]%
        {pretrain-rec}
\bibfield{author}{\bibinfo{person}{Zheni Zeng}, \bibinfo{person}{Chaojun Xiao},
  \bibinfo{person}{Yuan Yao}, {et~al\mbox{.}}}
  \bibinfo{year}{2021}\natexlab{b}.
\newblock \showarticletitle{Knowledge Transfer via Pre-training for
  Recommendation: A Review and Prospect}.
\newblock \bibinfo{journal}{\emph{Frontiers in Big Data}}
  (\bibinfo{year}{2021}).
\newblock


\bibitem[\protect\citeauthoryear{Zhang and Chen}{Zhang and Chen}{2020}]%
        {explain-rec-survey}
\bibfield{author}{\bibinfo{person}{Yongfeng Zhang} {and} \bibinfo{person}{Xu
  Chen}.} \bibinfo{year}{2020}\natexlab{}.
\newblock \showarticletitle{Explainable Recommendation: A Survey and New
  Perspectives}.
\newblock \bibinfo{journal}{\emph{Foundations and Trends in Information
  Retrieval}} (\bibinfo{year}{2020}).
\newblock


\bibitem[\protect\citeauthoryear{Zhang, Gu, Han, Chen, Xiao, Sun, Yao, Qi,
  Guan, Ke, Cai, Zeng, Tan, Liu, Huang, Han, Liu, Zhu, and Sun}{Zhang
  et~al\mbox{.}}{2021a}]%
        {cpm2}
\bibfield{author}{\bibinfo{person}{Zhengyan Zhang}, \bibinfo{person}{Yuxian
  Gu}, \bibinfo{person}{Xu Han}, {et~al\mbox{.}}}
  \bibinfo{year}{2021}\natexlab{a}.
\newblock \bibinfo{title}{CPM-2: Large-scale Cost-effective Pre-trained
  Language Models}.
\newblock
\newblock
\showeprint[arXiv]{2106.10715}


\bibitem[\protect\citeauthoryear{Zhang, Ma, Zhou, Men, Li, Ding, Tang, Zhou,
  and Yang}{Zhang et~al\mbox{.}}{2021b}]%
        {m6-ufc}
\bibfield{author}{\bibinfo{person}{Zhu Zhang}, \bibinfo{person}{Jianxin Ma},
  \bibinfo{person}{Chang Zhou}, {et~al\mbox{.}}}
  \bibinfo{year}{2021}\natexlab{b}.
\newblock \showarticletitle{M6-UFC: Unifying Multi-Modal Controls for
  Conditional Image Synthesis}. In \bibinfo{booktitle}{\emph{NeurIPS 2021}}.
\newblock


\bibitem[\protect\citeauthoryear{Zhou, Ma, Zhang, Zhou, and Yang}{Zhou
  et~al\mbox{.}}{2021}]%
        {clrec}
\bibfield{author}{\bibinfo{person}{Chang Zhou}, \bibinfo{person}{Jianxin Ma},
  \bibinfo{person}{Jianwei Zhang}, \bibinfo{person}{Jingren Zhou}, {and}
  \bibinfo{person}{Hongxia Yang}.} \bibinfo{year}{2021}\natexlab{}.
\newblock \showarticletitle{Contrastive Learning for Debiased Candidate
  Generation in Large-Scale Recommender Systems}. In
  \bibinfo{booktitle}{\emph{KDD 2021}}.
\newblock


\bibitem[\protect\citeauthoryear{Zhou, Song, Zhu, Fan, Zhu, Ma, Yan, Jin, Li,
  and Gai}{Zhou et~al\mbox{.}}{2018}]%
        {din}
\bibfield{author}{\bibinfo{person}{Guorui Zhou}, \bibinfo{person}{Chengru
  Song}, \bibinfo{person}{Xiaoqiang Zhu}, {et~al\mbox{.}}}
  \bibinfo{year}{2018}\natexlab{}.
\newblock \showarticletitle{Deep Interest Network for Click-Through Rate
  Prediction}. In \bibinfo{booktitle}{\emph{KDD 2018}}.
\newblock


\bibitem[\protect\citeauthoryear{Zhou, Zhao, Bian, Zhou, Wen, and Yu}{Zhou
  et~al\mbox{.}}{2020b}]%
        {kgsf}
\bibfield{author}{\bibinfo{person}{Kun Zhou}, \bibinfo{person}{Wayne~Xin Zhao},
  \bibinfo{person}{Shuqing Bian}, {et~al\mbox{.}}}
  \bibinfo{year}{2020}\natexlab{b}.
\newblock \showarticletitle{Improving Conversational Recommender Systems via
  Knowledge Graph based Semantic Fusion}. In \bibinfo{booktitle}{\emph{KDD}}.
\newblock


\bibitem[\protect\citeauthoryear{Zhou, Xu, Ge, McAuley, Xu, and Wei}{Zhou
  et~al\mbox{.}}{2020a}]%
        {bertpatience}
\bibfield{author}{\bibinfo{person}{Wangchunshu Zhou}, \bibinfo{person}{Canwen
  Xu}, \bibinfo{person}{Tao Ge}, {et~al\mbox{.}}}
  \bibinfo{year}{2020}\natexlab{a}.
\newblock \showarticletitle{BERT Loses Patience: Fast and Robust Inference with
  Early Exit}. In \bibinfo{booktitle}{\emph{NeurIPS 2020}}.
\newblock


\bibitem[\protect\citeauthoryear{Zhu and Gupta}{Zhu and Gupta}{2017}]%
        {gradualmagprune}
\bibfield{author}{\bibinfo{person}{Michael Zhu} {and} \bibinfo{person}{Suyog
  Gupta}.} \bibinfo{year}{2017}\natexlab{}.
\newblock \bibinfo{title}{To prune, or not to prune: exploring the efficacy of
  pruning for model compression}.
\newblock
\newblock
\showeprint[arXiv]{1710.01878}


\bibitem[\protect\citeauthoryear{Zou, Zhang, Cai, Ma, Cheng, Shi, Wang, Cheng,
  and Yin}{Zou et~al\mbox{.}}{2021}]%
        {ernie-rank}
\bibfield{author}{\bibinfo{person}{Lixin Zou}, \bibinfo{person}{Shengqiang
  Zhang}, \bibinfo{person}{Hengyi Cai}, {et~al\mbox{.}}}
  \bibinfo{year}{2021}\natexlab{}.
\newblock \showarticletitle{Pre-trained Language Model based Ranking in Baidu
  Search}. In \bibinfo{booktitle}{\emph{KDD 2021}}.
\newblock


\end{thebibliography}


    \clearpage
    \appendix
    \balance
    \section{Appendix}\label{sec:appendix}

\subsection{Datasets}\label{subsec:datasets}

We provide here the information on the datasets collected in our real-world systems, whose statistics are listed in Table~\ref{tab:dataset}.
Please refer to the related works for the information on the other datasets.

\paragraph{TaoProduct}
It is collected from the ranking stage of one of the large-scale product recommender systems of the mobile application Taobao, which serves hundreds of millions of users per day.
To avoid long training time, the negative samples, i.e., history records not clicked by users, are down-sampled by a factor of ten.
The training data and the validation data are collected from day one to day seven, while the test set is constructed from the data of day eight.
Infrequent items are filtered out, otherwise ID-based methods such as DIN will perform terribly on these cold items.

\paragraph{AlipayQuery}
It is collected from the ranking stage of the search query recommender system of the mobile application Alipay.
The negative samples are down-sampled by a factor of ten.
The training data and the validation data are collected from day one to day seven, while the test set is constructed from the data of day eight.
The user behaviors include the search queries and mini-apps that the users have previously interacted with, where infrequent queries and mini-apps are filtered out.

\paragraph{AliapyMiniApp}
It is collected from a retrieval stage responsible for retrieving mini-apps in the mobile application Alipay.
Mini-apps, also known as mini-programs, are lightweight apps with limited features that exist in a bigger main app that is Alipay in this case, and are developed by third-party developers instead of Alipay.
The training set and the validation set are constructed from data sampled from the past thirty days, and the test set is constructed from data sampled from the following seven days.
The number of items listed in Table~\ref{tab:dataset} counts both search queries and mini-apps, after filtering out rare or noisy items.
Due to the time gap between the training set and the test set, the test set contains a significant number of items not appeared in the training set.
For example, there are many novel search queries in the test set.

\begin{table}
    \centering
    \caption{
        Statistics of the datasets used for ranking and retrieval.
        M6-Rec uses only a small subset of the training data when experimenting with TaoProduct, AlipayQuery, and AlipayMiniApp due to the limited hardware resources available for training,
        even though the baselines such as DIN and YouTubeDNN do use all training data.
        Please refer to the related works for the information on the other tasks' datasets.
    }
    \label{tab:dataset}
    \begin{tabular}{@{}crrrr@{}}
        \toprule
        {\bf Dataset} & {\bf Train} & {\bf Valid} & {\bf Test} & {\bf \#Items}
        \\
        \midrule
        TaoProduct    & $\approx$500M & $\approx$200k & $\approx$200k & $\approx$30M \\
        AlipayQuery   & $\approx$200M & $\approx$200k & $\approx$200k & $\approx$5M  \\
        AlipayMiniApp & $\approx$300M & $\approx$200k & $\approx$200k & $\approx$80M \\
        Amazon-Movie  & 411,946       & 55,168        & 44,362        & 5,419        \\
        Amazon-Cloth  & 93,912        & 11,866        & 11,738        & 17,254       \\
        \bottomrule
    \end{tabular}
\end{table}

\subsection{Hyper-parameters}\label{subsec:hyperparameters}

The maximum sequence length is set to 1,024.
The low-rank adapters used by option-adapter tuning set the rank to $r=\frac{d}{H}$, where $d$ is the hidden size and $H$ is the number of attention heads.
The number of soft prompts is decided such that the total number of task-specific parameters account for roughly 1\% of the whole model's parameters unless otherwise specified, typically around 128.
The global batch size is 128.
We observe that fine-tuning and prompt-like tuning require different learning rates to achieve optimal performance.
In particular, prompt-like tuning necessitates a larger learning rate, which is at least the case for M6.
We thus use a learning rate of 0.00005 for fine-tuning, and use a learning rate of 0.0001 for prompt-like tuning including our option tuning and option-adapter tuning.

We train M6-Rec in a distributed manner on four machines, where each machine is equipped with eight Nvidia Tesla V100 GPUs.
Each training step of a dense deep model such as M6-Rec takes far more time than that of a sparse shallow model such as DIN and YouTubeDNN.
Therefore, M6-Rec uses only a small subset of the training data when running on TaoProduct, AlipayQuery, and AlipayMiniApp, even though the baselines such as DIN and YouTubeDNN do use all training data.
Specifically, we sample the training data such that a training session of M6-Rec lasts no more than eighteen hours.
The measurement of inference latency is also conducted on a V100 GPU, and the measurement includes the time spent on some miscellaneous overheads such as text preprocessing and embedding lookup.


\end{document}